\documentclass[]{article}

\usepackage[dvips]{graphicx}

\begin{document}

\title{Wilson's renormalization group applied to 2D lattice 
electrons in the presence of van Hove singularities}

\author{ B. Binz, D. Baeriswyl \\ {\small D\'epartement de Physique, Universit\'e de Fribourg, P\'erolles, CH-1700 Fribourg, Switzerland} \and B. Dou\c cot\\{\small Laboratoire de Physique de la Mati\`ere Condens\'ee, CNRS UMR 8551, Ecole Normale Sup\'erieure,}\\ {\small 24 rue Lhomond, 75231 Paris Cedex 05, France} \vspace{3mm} \\{\small Laboratoire de Physique Th\'eorique et Hautes Energies, CNRS UMR 7589, Universit\'es Paris VII,}\\ { \small 4 Place Jussieu, 75252 Paris Cedex 05, France}}

\newcommand{\be}{\begin{equation}}
\newcommand{\ee}{\end{equation}}
\newcommand{\ba}{\begin{eqnarray}}
\newcommand{\ea}{\end{eqnarray}}
\def\C{\mathbf{C}}
\def\R{\mathbf{R}}
\def\Z{\mathbf{Z}}
\renewcommand{\v}[1]{{\bf #1}}
\def\mod{\mathop{\rm mod}}
\def\arsinh{\mathop{\rm arsinh}}
\def\up{\uparrow}
\def\down{\downarrow}
\newcommand{\ud}{\mathrm{d}}
\def\Min{\mathop{\rm Min}}
\def\Max{\mathop{\rm Max}}
\def\P{\mathop{\rm P}}
\def\sign{\mathrm{sign}}

\def\Map#1{\smash{\mathop{\hbox to 35 pt{$\,$\rightarrowfill$\,\,$}}
                             \limits^{\scriptstyle#1}}}
\def\MapRight#1{\smash{\mathop{\hbox to 35pt{\rightarrowfill}}
\limits^{#1}}}

\maketitle

\abstract{The weak coupling instabilities of a two dimensional 
Fermi system are investigated for the case of a square
lattice using a Wilson renormalization group
scheme to one loop order. 
We focus on a situation where the Fermi surface passes
through two saddle points of the single particle dispersion. In the
case of perfect nesting, the dominant instability is a spin density
wave but $d$-wave superconductivity as well as charge or spin flux
phases are also obtained in certain regions in the space of coupling 
parameters. The low energy regime in the
vicinity of these instabilities can be studied analytically.  Although
saddle points play a major role (through their large contribution to
the single particle density of states), the presence of low energy
excitations along the Fermi surface rather than at isolated points is
crucial and leads to an asymptotic decoupling of the various
instabilities. This suggests a more mean-field like  picture
of these instabilities, than the one recently established by numerical
studies using discretized Fermi surfaces.}


\section{Introduction}

Most of the unusual properties of the superconducting cuprates are 
likely to be linked to the quasi-two-dimensional nature of their 
electronic structure close to the Fermi energy. Therefore certain 
(single-band) two-dimensional (2D) models of interacting electrons may
 be able, in principle, to account for at least part of the 
anomalies observed in these compounds 
\cite{Anderson,ZhangRice,Orenstein}. Unfortunately, even very
 simple models, such as the 2D Hubbard or the 2D $t-J$ model, 
have so far resisted a rigorous analysis. Moreover, the available
 numerical studies are not yet conclusive enough for making 
definite predictions for, e.g., the zero-temperature phase diagram
 of these many-electron systems. 

One of the major difficulties is that in the cuprates the bare 
couplings between electrons, for instance the parameter $U$ of 
the Hubbard model, are large, i.e. of the order of the bandwidth.
 Therefore it is not clear whether a ground state consisting of
 occupied Bloch orbitals with energies below $\epsilon_F$ is a good 
starting point or whether one has rather to think in terms of 
configurations of singly occupied and empty sites (doped Mott 
insulator). Actually, the successful analysis of the insulating
 phase in terms of the Heisenberg model suggests that the Mott 
insulator is the appropriate reference state \cite{Manousakis}. 
Another difficulty is that fluctuations (both thermal and 
quantum) are strong in two dimensions so that mean-field 
approximations cannot be trusted.

In this paper we deliberately choose the limit of weak bare
 couplings, keeping in mind that this parameter range may 
miss completely some important characteristic aspects of 
the region of strong bare interactions. Nevertheless, it 
cannot be excluded that certain properties are qualitatively
 the same over the whole range of (bare) couplings, as is 
the case for the 1D Hubbard model, a Luttinger liquid for 
all positive values of $U$ and all densities except $n=1$ 
\cite{Shiba,Voit}.

The most clear picture of two-dimensional interacting 
electrons has been obtained for the 2D jellium model 
with its circular Fermi surface, using a Wilsonian 
renormalization group (RG) approach \cite{Feldman,Intro,Dupuis}.
 A series of rigorous studies has shown that the Landau 
Fermi liquid theory is stable at not too low temperatures
 \cite{Salmhofer,Disertori}, i.e., above the critical 
temperature for Kohn-Luttinger superconductivity 
\cite{Kohn}. Other instabilities do not occur.

Electrons hopping between the sites of a square 
lattice yield a spectrum that differs in two 
respects from the parabolic spectrum of the 
jellium model. First, the spectrum exhibits 
extrema and saddle points in the Brillouin 
zone. General considerations imply that there
 are at least two saddle points and two extrema
 (one maximum and one minimum). Obvious points
 are $\v P_0=(0,0)$ and $\v Q=(\pi,\pi)$ for the
 extrema and $\v P_1=(\pi,0)$ and $\v P_2=(0,\pi)$
 for the two saddle points, but more complicated 
patterns are also possible. The density of states
 has a logarithmic van Hove singularity at the 
saddle points, in strong contrast to the constant
 density of states of the parabolic spectrum of 
the jellium model. The second difference is the 
curvature of the lines of constant energy. These
 are circles in the case of the jellium model, 
whereas in the case of the square lattice one 
can easily find portions with almost vanishing
 curvature. In fact, for the tight-binding model
 (with hopping restricted to nearest neighbor 
sites) the Fermi surface for the half-filled band
 is a perfect square.

RG calculations for a model where the Fermi 
surface contains flat portions have
 been performed by various authors \cite{Fernao,
Kwon,ZYD,Madrid-inf}. They agree in that a 
$d$-wave superconducting instability occurs for 
repulsive interactions, due to the coupling
 of particle-particle and particle-hole correlations. 

Our main emphasis is on the effect of van Hove
 singularities. We will consider in particular
 the case where the Fermi surface passes through
 saddle points (``van Hove filling''). 
Early scaling approaches to this problem 
\cite{Schulz1,Dzyalo1,Lederer} focussed on the 
interactions between electrons at the saddle
points, by treating these points in analogy to 
the two Fermi points of the one-dimensional 
electron gas \cite{Solyom}. In this work we show that, indeed,
 the logarithmically dominant RG flow at low
 energies is controlled by the neighborhood of the
 van Hove points. However, in contrast to the one-dimensional case where the 
scattering processes can be characterized in terms of a few 
coupling {\it 
constants } 
connecting the two Fermi points, in two 
dimensions the effective couplings are {\it 
functions } 
of incoming 
and outgoing momenta, even if these are restricted to the 
Fermi surface. We find that this functional dependence plays 
 a crucial role in the asymptotic decoupling of competing 
 instabilities. A step in this direction has already been made in 
the parquet approach of Ref. \cite{Yakovenko}.

When the Fermi level is at a van Hove 
singularity the system is not renormalizable
 in the traditional sense of field theory. 
Nevertheless, electrons near a van Hove singularity 
have been treated by applying the field theory formalism 
 to the particle-hole sector \cite{Madrid-vH}. No mixture 
with particle-particle diagrams can be treated 
within this formalism. The Wilsonian
 RG used here does not assume renormalizability and may be 
applied without constraints. 

A numerical scheme for calculating the complete flow from the bare 
action of an arbitrary microscopic model to the low-energy effective
 action as a function of a continuously decreasing energy cutoff 
$\Lambda$ has been presented by Zanchi and Schulz \cite{Drazen}. 
Unfortunately, in order to carry out the RG calculations it appears 
to be necessary to resort to a number of approximations, which are 
justified only at the final stage of the RG flow, where $\Lambda$ 
is much smaller than the bandwidth. Nevertheless, the application 
of this method to the Hubbard model near half filling does provide 
an appealing picture, namely a transition from an antiferromagnetically 
ordered ground state at half filling to a $d$-wave superconductor upon 
doping \cite{Drazen}. This result has been confirmed by Halboth and 
Metzner using a similar approach \cite{Halboth}. Recently, numerical
 RG calculations have brought up two additional phases, one with a 
deformed Fermi surface (Pomeranchuk instability) \cite{Halboth2} 
and one with suppressed uniform spin and charge susceptibilities 
(``insulating spin liquid'') \cite{Honerkamp}. In all these 
calculations the proximity of van Hove singularities plays an 
important role, together with approximate nesting.

Our analytical approach is complementary to these numerical RG 
calculations. We start from the same equations and analyze the flow in the limit of small $\Lambda$. We focus on the system at the van Hove 
filling, where we take only the leading order in $\Lambda$ into account. The 
asymptotic regime of small $\Lambda$ can only be reached, if the initial coupling is 
sufficiently small. In this sense our approach is limited as compared
to the numerical studies. On the other hand, the numerical methods
suffer from the need to replace the continuous Fermi surface by a
discrete set of points. 

 In our approach - as well as in previous RG calculations carried out to one loop order - self-energy effects are neglected. While this can be easily justified for the jellium model, the argument is more subtle in the case of lattice fermions. In fact, the second-order contribution to the self-energy is infrared divergent in the case of the half-filled nearest-neighbor tight-binding band. Nevertheless, we find that also in this case self-energy effects are of subleading order in $\Lambda$, provided that an instability (superconducting or density wave) occurs.

The analysis of the dominant parts of the RG equations is sufficient 
for establishing a rich phase diagram for the nearest-neighbor tight-binding band with 
a nested Fermi surface, whereas in the non-nested case subleading contributions are crucial. We also point out the difficulties of including consistently 
those subleading terms.

The paper is organized as follows. In Section \ref{2} we define the effective 
interaction for a general model of interacting electrons in 
two dimensions and show how to relate it to
  correlation functions. An exact RG equation and its one loop 
approximation is presented in Section \ref{rgsection} both for the effective coupling function and for generalized susceptibilities. The close connection between this approximation and parquet diagrams is also briefly discussed.  In Section \ref{lead},
 we analyze the one loop equations in the limit of small energies. First, we 
review the situation of a parabolic electron dispersion with a circular Fermi
 surface where we recover the standard result of a dominant flow
  in the BCS channel. Then a Fermi surface is considered which passes through van Hove points without being nested. For attractive interaction superconductivity again dominates, whereas it appears to be difficult to keep track consistently of all leading order terms for repulsive interaction.
 The case of a half filled nearest-neighbor  tight-binding band is discussed in the remaining Sections \ref{squareFS}-\ref{symmsection}. 
In Section \ref{squareFS}, the renormalized couplings are classified according to both the location of momenta with respect to the van Hove points and the channels characterizing the different instabilities. It is argued that to leading order there is no mixing between superconducting, charge and spin instabilities except from momenta very close to the van Hove points. A simple way of disentangling this special behavior at the van Hove points and the generic behavior elsewhere is presented in Section \ref{2patch} and contrasted to an earlier approach where the momentum dependence was altogether neglected. The asymptotic behavior of the RG flow  allows to draw a phase diagram including superconductivity, density waves and flux phases, depending on the values of the bare couplings. This phase diagram agrees with symmetry considerations linking the various order  parameters, as shown in Section \ref{symmsection}. A brief summary is presented in Section \ref{conclusion}.

\section{Effective interaction}\label{2}

We consider a system of interacting electrons on a two-dimensional square 
lattice.
 The single particle states are labeled  by a lattice momentum 
$\v k$, which is defined only modulo a reciprocal lattice vector, and a spin index $\sigma=\up,\down$. In the 
functional 
integral formalism the  system is described in terms of Grassmann fields 
$\psi_{\sigma k}$, where $k=(k_0,\v k)$ is a
 $2+1$-dimensional variable, which includes the Matsubara frequency $k_0$ \cite{NegeleOrland}. The Fourier transform is defined as 
$$\psi_{\sigma k}=(\beta V)^{-1/2}\int_0^\beta\ud\tau\, e^{ik_0\tau}\sum_{\v r}e^{-i\v k\v r}\psi_{\sigma\v r}(\tau),$$ 
where $\beta$ is the inverse temperature and $V$ the volume of the system. In the calculations we will take the limit $\beta,V\to\infty$. The action is of the form 
 \be S[\psi]=\sum_{\sigma,k}\bar\psi_{\sigma k}(ik_0-\xi_{\v k}) \psi_{\sigma k}-W[\psi],
\label{S}\ee
where $\xi_{\v k}=e_{\v k}-\mu$ is the single particle energy relative to 
the chemical potential. The free electron propagator is 
\be C(k)=\frac 1{ik_0-\xi_{\v k}}.\label{C}\ee
$W[\psi]$ can be a general short ranged two-body interaction, with 
a coupling function $g(k_1, k_2, k_3, k_4)$,
\ba 
 W[\psi]&=&\frac12\frac1{\beta V}\sum_{k_1\cdots k_4}
\delta_{k_1+k_2,k_3+k_4}\nonumber\\
& &\quad\times g(k_1, k_2, k_3, k_4) \sum_{\sigma,\sigma'}
\bar\psi_{\sigma k_1}\bar\psi_{\sigma' k_2}\psi_{\sigma' k_3}\psi_{\sigma k_4}.
\label{int}
\ea
Note that although we write formally $\delta_{k_1+k_2,k_3+k_4}$, the
momenta are only conserved modulo a reciprocal lattice vector.

The function $g$ should satisfy all the point symmetries of the square 
lattice. In addition, we require permutation symmetry 
$g(k_1,k_2,k_3,k_4)=g(k_2,k_1,k_4,k_3)$. Finally, from time 
reversal symmetry and the behavior under complex conjugation one finds 
\cite{Halboth} $g(k_1,k_2,k_3,k_4)=g(k_4,k_3,k_2,k_1)$ 
$=\bar g(\bar k_1,\bar k_2,\bar k_3,\bar k_4)$, where $\bar k=(-k_0,\v k)$ and 
$\bar g$ is the complex conjugate of $g$.

There is no symmetry with respect to the operation 
$Xg(k_1,k_2,k_3,k_4):=g(k_1,k_2,k_4,k_3)$ which
exchanges only two of its arguments, but $g$ can be separated into a
symmetric part  $g^{S}=\frac12(1+X)g$ and an
antisymmetric part $g^{T}=\frac12(1-X)g$. These couplings describe scattering 
of singlet and triplet pairs, as becomes clear if we write the interaction as
\ba 
 W[\psi]&=&\frac12\frac1{\beta V}\sum_{k_1\cdots k_4}
\delta_{k_1+k_2,k_3+k_4}\ \bigg\{ g^S(k_1,k_2,k_3,k_4)
\bar\phi_S(k_2,k_1)\phi_S(k_3,k_4)\nonumber\\
& &\qquad + g^T(k_1,k_2,k_3,k_4)
\sum_{\alpha=0,\pm1}\bar\phi_{\alpha}(k_2,k_1)\phi_{\alpha}(k_3,k_4)\bigg\},
\ea
with 
\ba
&\phi_{S}(k,k')=\frac1{\sqrt{2}}\left(\psi_{k\up }\psi_{k'\down }-\psi_{k\down }
\psi_{k'\up}
\right).&\\
&\phi_{0}(k,k')=\frac1{\sqrt{2}}\left(\psi_{k\up }\psi_{k'\down }+\psi_{k\down }
\psi_{k'\up}
\right)\ ,\ 
\phi_{1}(k,k')=\psi_{k\up }\psi_{k'\up }\ ,\
\phi_{-1}(k,k')=\psi_{k\down }\psi_{k'\down }&\nonumber
\ea

The system is completely described  by the partition function with source term
\be 
Z[\eta]=\int d\mu_C[\psi]\,  e^{-W[\psi]+(\bar\eta,\psi)+(\bar\psi,\eta)},
\label{Z}
\ee
where we used the short-hand notation 
 $(\bar\chi,\psi):=\sum_{\sigma k} \bar\chi_{\sigma k}\psi_{\sigma k}$ and the normalized
Gaussian measure is  defined by
\be 
d\mu_C[\psi]:=\frac{\prod_{\sigma k}d\psi_{\sigma k}
d\bar\psi_{\sigma k}\ e^{(\bar\psi,C^{-1}\psi)}}{\int\prod_{\sigma k}d\psi_{\sigma k}
d\bar\psi_{\sigma k}\ e^{(\bar\psi,C^{-1}\psi)}}.
\ee
In particular, all connected correlation functions are obtained as 
functional derivatives \cite{NegeleOrland}
\be
\langle\psi_1\cdots\psi_n\bar\psi_{n+1}\cdots\bar\psi_{2n}\rangle_{\rm c}=
\left.\frac{\delta^{2n}\log Z[\eta]}{\delta\!\eta_{2n}\cdots
\delta\!\eta_{n+1}
\delta\!\bar\eta_{n}\cdots\delta\!\bar\eta_{1}}\right|_{\eta=0},
\ee
where we have written $\psi_i$ instead of $\psi_{k_i\sigma_i}$.

The propagator (\ref{C}) is singular on the manifold $k_0=\xi_{\v k}=0$
 called the Fermi surface. The infinities encountered in a naive perturbative 
treatment of the action
(\ref{S}) are avoided in the RG procedure which endows the bare propagator 
with an infrared cutoff $\Lambda$, 
\be 
C_\Lambda(k)=\Theta(|\xi_{\v k}|-\Lambda)C(k),\label{sharp}
\ee
where $\Theta$ is the Heavyside step function. The quantity that separates high- 
from low-energy degrees of freedom is therefore $|\xi_{\v k}|$. Although a 
more canonical choice would be $\sqrt{k_0^2+\xi^2(\v p)}$, we have chosen the
 frequency-independent cutoff in order to simplify the calculations.

One can now define the {\it 
effective interaction }
\be
{\cal W}_\Lambda[\chi]=-\log  \int d\mu_{C_\Lambda}[\psi]
e^{-W[\psi+\chi]}\label{effint}
\ee
which depends on a Grassmann field $\chi$.
Note that the integration with respect to $d\mu_{C_\Lambda}[\psi]$ is perfectly
 defined, 
although $C_\Lambda^{-1}$ is not. This can be seen most easily in the 
expansion of ${\cal W}_\Lambda[\chi]$ in terms 
of Feynman diagrams. The evaluation of these diagrams involves only 
$C_\Lambda$ and never $C_\Lambda^{-1}$. Whenever  $C_\Lambda^{-1}$ 
appears in an intermediate step of a calculation (see below), it may be 
regularized by replacing the zero in the Heavyside function by an 
infinitesimal number.

${\cal W}_\Lambda$  has a twofold interpretation. On the one hand, we 
can restrict the field to the low energy degrees of freedom 
$\psi_{{\rm <}\, k\sigma}=\Theta(\Lambda-|\xi_{\v k}|)\psi_{k\sigma}$. 
The object
\be
S^{\rm eff}_\Lambda[\psi_{\rm <}]=(\bar\psi_{\rm <},C^{-1} \psi_{\rm <})-
{\cal W}_\Lambda
[\psi_{\rm <}]
\ee
corresponds then to Wilson's effective action, which describes the system in 
terms of $\psi^{\rm <}$ only.

On the other hand, ${\cal W}_\Lambda$ is the generating functional of amputated
 connected correlation functions with infrared cutoff $\Lambda$ because of the
 identity \cite{Salmhoferbook} 
\be
\log Z_\Lambda[\eta]=-(\bar\eta,C_\Lambda \eta)-{\cal W}_\Lambda[C_\Lambda\eta]
,\label{gen}
\ee
where $Z_\Lambda$ is given by (\ref{Z}), with $C$ replaced by 
$C_\Lambda$. In particular,  the quadratic part of ${\cal W}_\Lambda$ is related
 to the self-energy $\Sigma_\Lambda$ by 
\ba
\left.\frac{\delta^2}{\delta\!\chi_{\sigma k}\delta\!\bar\chi_{\sigma k}}
{\cal W}_\Lambda[\chi]\right|_{\chi=0}&=&-C^{-1}_\Lambda(k)-\langle
\psi_{\sigma k}\bar\psi_{\sigma k}\rangle_{\rm \Lambda}\,C^{-2}_\Lambda(k)\\
&=&\frac{\Sigma_\Lambda(k)}{1-C_\Lambda(k)\Sigma_\Lambda(k)},\label{self}
\ea
where we have used the following identity for the full electron propagator 
$G_\Lambda(k)=-\langle\psi_{\sigma k}\bar\psi_{\sigma k}\rangle_{\rm \Lambda}
=C_\Lambda(k)(1-C_\Lambda(k)\Sigma_\Lambda(k))^{-1}$. 
Therefore in the case $|\xi_{\v k}|<\Lambda$ the right-hand side of Eq. 
(\ref{self}) simply becomes $\Sigma_\Lambda(k)$. 

Similarly the quartic part of  ${\cal W}_\Lambda$ is related to the one particle 
irreducible vertex $\Gamma_\Lambda$, defined by  $\langle\psi_{\sigma k_1}
\psi_{\sigma' k_2}\bar\psi_{\sigma' k_3}\bar
\psi_{\sigma k_4}\rangle_{c,\Lambda}=(\beta V)^{-1}\Gamma_\Lambda^{\sigma\sigma'}(k_1,\ldots,k_4)
\,\prod_{i=1}^4G_\Lambda(k_i)$. In fact,
differentiating  Eq. (\ref{gen}) we find 
\ba
\left.\frac{\delta^{4}{\cal W}_\Lambda[\chi]}
{\delta\!\chi_{\sigma k_4}\delta\!\chi_{\sigma' k_3}\delta\!\bar
\chi_{\sigma' k_2}\delta\!\bar\chi_{\sigma k_1}}\right|_{\chi=0}
&=&-\langle\psi_{\sigma k_1}\psi_{\sigma' k_2}\bar\psi_{\sigma' k_3}\bar
\psi_{\sigma k_4}\rangle_{\rm c,\Lambda}\prod_{i=1}^4C_\Lambda^{-1}(k_i)\\
&=&-\frac{\Gamma_\Lambda^{\sigma\sigma'}(k_1,\ldots,k_4)}{\beta V\prod_{i=1}^4
[1-C_\Lambda(k_i)\Sigma_\Lambda(k_i)]}.\label{vert}
\ea

The quartic part of ${\cal W}_\Lambda$ is of the same form as (\ref{int}) 
with an effective coupling function $g_\Lambda(k_1,\ldots,k_4)$. For  $|\xi_{\v k_i}|<\Lambda$ we find therefore
\ba
\Gamma^{\sigma \sigma'}_{\Lambda}(k_1,\ldots,k_4)
&=&-\,\delta_{k_1+k_2,k_3+k_4}\,(1-\delta_{\sigma \sigma'}X)
g_\Lambda(k_1,\ldots,k_4)\label{vertex}.
\ea
$g_\Lambda$ is equal to a connected amputated 
correlation function if all $|\xi_{\v k_i}|>\Lambda$ and a one particle 
irreducible vertex in the opposite case. $g_\Lambda$ is therefore not continuous
at $|\xi_{\v k_i}|=\Lambda$.
A formal and non-perturbative proof of these relations was given  by Morris 
\cite{Morris} for a bosonic field theory but the generalization to fermions 
is straightforward. He has also shown that $\Sigma_\Lambda$ and 
$\Gamma_\Lambda$ are continuous at $|\xi_{\v k_i}|=\Lambda$, in contrast to $g_\Lambda$.

\section{RG equations}\label{rgsection}

\subsection{RG flow of the effective interaction}

The effective interaction satisfies the following exact RG equation
\cite{Salmhofer,Salmhoferbook} 
\be 
\frac\ud{\ud\Lambda}{\cal W}_\Lambda[\chi]=\sum_{\sigma,k}\frac{\ud C_\Lambda}{\ud \Lambda}(k)\
\frac{\delta^2{\cal W}_\Lambda[\chi]}{\delta\!\chi_{\sigma k}\delta\!
\bar\chi_{\sigma k}}-\sum_{\sigma,k}\frac{\ud C_\Lambda}{\ud \Lambda}(k)\ \frac{\delta{
\cal W}_\Lambda[\chi]}{\delta\!\chi_{\sigma k}}\frac{\delta{
\cal W}_\Lambda[\chi]}{\delta\!\bar\chi_{\sigma k}}.\label{exact}
\ee 
 The derivative in $\frac\ud{\ud\Lambda}C_\Lambda$ restricts the propagator to an infinitesimal energy shell $(\Lambda,\Lambda+\ud\Lambda)$ and 
Eq. (\ref{exact}) therefore describes the effect of integrating out the 
 modes of that energy shell.
This equation was first derived by Polchinski in the context of a scalar field theory \cite{Polchinski}.

The strategy is now to start with the bare interaction (\ref{int}) at a cutoff 
$\Lambda_0$ and to use some truncation of Eq. (\ref{exact}) in order to compute
 approximately the effective action in the limit $\Lambda\to0$. We will follow here Zanchi and Schulz \cite{Drazen} who proposed to develop ${\cal W}_\Lambda$ up to order six in the fermionic variables and to neglect terms of higher order. 

 We will only follow the flow of the effective coupling function and neglect self energy corrections (i.e. the quadratic part of ${\cal W}_\Lambda$). Self energy corrections to the single particle Green's function have three main effects. First they change the shape and location of the Fermi surface, second they modify the properties of the single particle dispersion (the Fermi velocity) and third they lead to a reduction of the quasi particle weight. It has to be checked from case to case whether such corrections can be safely neglected or not (see Sections \ref{circ}, \ref{vHove} and \ref{squareFS}).

Terms of order six are not present in the original interaction but they are 
produced by the RG flow.
Their effect is then to renormalize the effective coupling function $g_\Lambda$.
 The result is a closed one-loop equation for the coupling function 
$g_\Lambda(k_1,\ldots,k_4)$ \cite{Drazen} where $|\xi_{\v k_i}|<\Lambda$. It 
reads 
\be 
\frac\ud{\ud\Lambda}g_\Lambda(k_1,\ldots,k_4)=\mbox{PP}+\mbox{PH1}+\mbox{PH2}
\label{1loop}
\ee
$$ 
\mbox{PP}=-\frac1{\beta V}\sum_p\frac{\ud\left[C_\Lambda(p)C_\Lambda(q_0)\right]}{\ud\Lambda}
g_{\tilde\Lambda_0}(k_1,k_2,q_0,p) g_{\tilde\Lambda_0}(p,q_0,k_3,k_4), 
$$
$$ 
\mbox{PH1}=-\frac1{\beta V}\sum_p\frac{\ud\left[C_\Lambda(p)C_\Lambda(q_1)\right]}{\ud\Lambda}
g_{\tilde\Lambda_1}(k_1,q_1,k_3,p) g_{\tilde\Lambda_1}(p,k_2,q_1,k_4), 
$$
\begin{eqnarray*}
\mbox{PH2}&=&-\frac1{\beta V}\sum_p\frac{\ud\left[C_\Lambda(p)C_\Lambda(q_2)\right]}{\ud\Lambda}
\left[-2g_{\tilde\Lambda_2}(k_1,p,q_2,k_4) g_{\tilde\Lambda_2}(q_2,k_2,k_3,p)\right.\\& 
 &\qquad\left.+g_{\tilde\Lambda_2}(k_1,p,k_4,q_2) g_{\tilde\Lambda_2}(q_2,k_2,k_3,p)+
g_{\tilde\Lambda_2}(k_1,p,q_2,k_4) g_{\tilde\Lambda_2}(q_2,k_2,p,k_3)\right],
\end{eqnarray*}
 where  $q_0=k_1+k_2-p$, $q_1=p+k_3-k_1$, $q_2=p+k_3-k_2$ and $\tilde\Lambda_i=\Max\{|\xi_{\v p}|,|\xi_{\v q_i}|\}$.

\begin{figure}  
\centerline{\includegraphics[width=11cm]{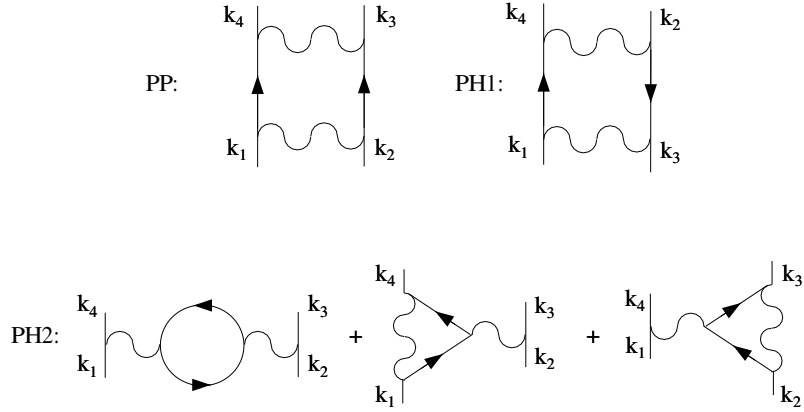}}
\caption{Diagrams contributing to the renormalization of the coupling
function $g_\Lambda(k_1,\ldots,k_4)$.}\label{diags}
\end{figure}

Note that unlike Eq. (\ref{exact}) which is valid for any choice of cutoff 
function $C_\Lambda$, this truncated equation assumes a sharp cutoff as in Eq.
 (\ref{sharp}).
Its representation in terms of Feynman diagrams is shown in Fig. \ref{diags}. 

One of the two internal lines stands for a propagator $C_\Lambda$ and 
the other for its derivative $\frac\ud{\ud\Lambda}C_\Lambda$. This amounts to a loop 
integration over an energy shell 
$\Lambda\leq|\xi_{\v p}|\leq\Lambda+\ud\Lambda$ 
with the restriction that the second propagator is in the high energy regime 
$|\xi_{\v q}|\geq \Lambda$. The wavy lines  in the diagrams represent 
effective couplings $g_{\tilde\Lambda}$ of an earlier stage of the RG flow,
 when the energy shell $\tilde\Lambda$ was integrated out. 

Eq. (\ref{1loop}) is not local in the variable $\Lambda$. Since this is not very 
convenient, it was proposed \cite{Halboth} 
to develop Eq. (\ref{exact}) into Wick ordered polynomials of the fermionic 
variables instead of monomials as it was done above. Wick ordering with
 respect to the low energy propagator $D_\Lambda=C-C_\Lambda$ 
results in the same one-loop equation as above but now all the couplings are 
evaluated at the actual RG variable $\Lambda$ and $-\ud\left[C_\Lambda(p)
C_\Lambda(q)\right]/{\ud\Lambda}$ has to be replaced by $\ud\left[D_\Lambda(p)
D_\Lambda(q)\right]/{\ud\Lambda}$, i.e., the energy of the second propagator is now restricted to be {\it 
smaller} 
than $\Lambda$. 

Recently a third scheme was established \cite{Honerkamp} where the equation 
is  local in $\Lambda$ and the second propagator is still in the high energy 
sector. It is obtained by considering the generating functional of the
one particle irreducible vertex functions rather than the effective
interaction defined by Eq. (\ref{effint}), an idea originally due to
Wetterich \cite{Wetterich}.

\subsection{Order parameters and generalized susceptibilities}\label{corr}

The RG formalism can be used to calculate the linear response 
to an external field \cite{Drazen,Halboth}. We focus on the prominent 
instabilities of the half-filled nearest-neighbor tight-binding band by adding 
the term 
\be 
W'[h,\psi]=-\bar h\int_0^\beta\!d\tau\,O(\tau)-h\int_0^\beta\!d\tau\,\bar O(\tau),\label{pert}
\ee
to the action, where h is the 
external field $h\in\C$  and $O$ is a characteristic order parameter. 

As a first example we consider superconductivity, described 
 by the order parameter $O=\sum_{\v k}f(\v k)\,\psi_{\up \v k}\psi_{\down -\v k}$ with a form factor $f(\v k)$ 
 depending on the symmetry chosen (s-wave, d-wave, etc.). The 
 effective interaction, defined by Eq. (\ref{effint}) with $W$ replaced 
by $W+W'$, is now 
\be
{\cal W}_\Lambda[h,\psi]=\left.{\cal W}_\Lambda[\psi]\right|_{h=0}-\left(\bar h\sum_kR_\Lambda(k)\psi_{\up k}\psi_{\down -k}+\mbox{h.c.}\right)-\chi_\Lambda\,\beta V\, |h|^2\label{terms}
\ee
plus terms of higher order in $h$ and $\psi$. 
 The coefficients $R_\Lambda(k)$ and $\chi_\Lambda$ are actually defined by Eq. (\ref{terms}). Their initial values are $R_{\Lambda_0}(k)=f(\v k)$ and $\chi_{\Lambda_0}=0$.  The susceptibility $\chi_\Lambda$ gives the linear response with respect to the perturbation (\ref{pert}) of the system with an infrared cutoff $\Lambda$.
Using Eq. (\ref{terms}) and the definition (\ref{effint}) we get
\be
\chi_\Lambda=-\frac1{\beta V}\left.\frac{\partial^2{\cal W}_\Lambda[h,\psi]}{\partial h\partial\bar h}\right|_{h,\psi=0}=\frac1{V}\left.\frac{\partial\langle O\rangle_\Lambda}{\partial h}\right|_{h=0}.
\ee 
$\chi_\Lambda$ is therefore interpreted as a susceptibility at temperature $T=\Lambda$ \cite{Drazen}. In fact, in finite temperature perturbation theory the temperature acts like an infrared cutoff. 

One can now derive one loop RG equations for $R_\Lambda(k)$ and $\chi_\Lambda$ following the same procedure as for the effective coupling function $g_\Lambda$. We find
\ba
\frac\ud{\ud\Lambda}\chi_\Lambda&=&\frac1{\beta V}\sum_k\frac\ud{\ud\Lambda}\left[C_\Lambda(k)C_\Lambda(-k)\right]\,|R_\Lambda(k)|^2,\nonumber\\
\frac\ud{\ud\Lambda}R_\Lambda(k)&=&-\frac1{\beta V}\sum_p\frac\ud{\ud\Lambda}\left[C_\Lambda(p)C_\Lambda(-p)\right]\,g_\Lambda^{BCS}(k,p)\,R_\Lambda(p),\label{ppresponse}
\ea
where the BCS coupling function $g_\Lambda^{BCS}(
k, k'):=g_\Lambda( k,- k,- k', k')$ describes the
scattering of Cooper pairs with zero total momentum. Note that the function $R_\Lambda(\v k)$ has the same symmetry properties ($s$-wave, $d$-wave etc.) with respect to the lattice point group as the initial form factor $f(\v k)$. 

At half filling,
instabilities related to the nesting
vector $\v Q=(\pi,\pi)$ are also natural candidates. They are described by the order parameter 
\be
O=\sum_{\v k,\sigma}f_\sigma(\v k)\,\bar\psi_{\sigma \v k+\v Q}\,\psi_{\sigma \v k}=\sum_{\v r,\v r',\sigma}e^{i\v r\v Q}\,\tilde f_\sigma(\v r-\v r')\,\bar\psi_{\sigma\v r}\psi_{\sigma\v r'},\label{orderparameter}
\ee
where $\tilde f$ is the Fourier transform of $f$. 
We assume that $f_\sigma(\v k+\v Q)=\bar f_\sigma(\v k)$ so that $O$ is Hermitian and $h\in\R$.

\begin{figure} 
        \centerline{\includegraphics[width=4cm]{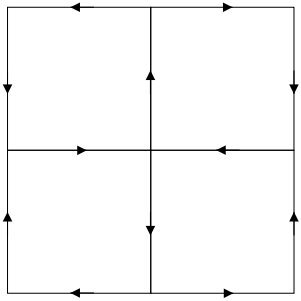}}
\caption{The pattern of charge (spin) currents along the bonds of 
the square lattice in a charge (spin) flux phase}\label{flux}
\end{figure}

 The simple choices $f_\sigma(\v k)=1$ and $f_\sigma(\v k)=\sigma$ yield, respectively, charge and spin density waves. If
 $f_\sigma(\v k)$ is a function with $d_{x^2-y^2}$-symmetry and is even
 (odd) in the spin index, then the nearest-neighbor-terms  in Eq. (\ref{orderparameter}) yield circular charge (spin) currents flowing around
 the plaquettes of the square lattice with alternating directions (see 
Fig. \ref{flux}). These four charge and spin instabilities have been discussed a long time ago in the context of the excitonic insulator \cite{Halperin}.

We call the phase with circulating charge currents the charge flux phase (CF) \cite{flux}, it is sometimes also called $d$-density wave, charge current wave or orbital antiferromagnetism. The charge flux phase (CF), closely related to the concept of the chiral spin liquid
\cite{chiral}, still plays a prominent role in the strong
 coupling $SU(2)$ theory of the $t-J$ model \cite{su2}. Recently it
 was proposed to be {\it the} competing order parameter to $d$-wave
 superconductivity and responsible for the pseudo gap phase of the 
cuprates \cite{Laughlin}. 

The phase with circulating spin currents is called the spin flux phase (SF). Other names encountered in the literature are ``spin current wave'' or ``spin nematic state'' (because it is a state with broken rotational symmetry and unbroken time reversal symmetry).
The low-temperature thermodynamics of both the charge and spin flux phases have been investigated within a mean field theory in Ref. \cite{Nersesyan}.

The effective interaction in the presence of such a perturbation is of the form
\be
{\cal W}_\Lambda(h)[\psi]={\cal W}_\Lambda(0)[\psi]-2h\sum_{k,\sigma}R_{\sigma\Lambda}(k)\bar\psi_{\sigma k+Q}\,\psi_{\sigma k}-\chi_\Lambda\,\beta V\, h^2+\ldots,
\ee
where $Q:=(0,\v Q)=(0,\pi,\pi)$. The one loop equations involve the scattering processes $g^d(k,k'):=Xg^x(k,k'):=g(k,k'+Q,k',k+Q)$  with a direct or
exchanged momentum transfer of $\v Q$. They read
\ba
\frac\ud{\ud\Lambda}\chi_\Lambda&=&-2\frac1{\beta V}\sum_{p,\sigma}\frac\ud{\ud\Lambda}\left[C_\Lambda(p)C_\Lambda(p+Q)\right]\,|R_{\sigma\tilde\Lambda}(p)|^2,\label{phresp}\\
\frac\ud{\ud\Lambda}R_{\sigma\Lambda}(k)&=&\frac1{\beta V}\sum_p\frac\ud{\ud\Lambda}\left[C_\Lambda(p)C_\Lambda(p+Q)\right]\sum_{\sigma'}\left\{g_{\tilde\Lambda}^{d}(k,p)-\delta_{\sigma\sigma'}g_{\tilde\Lambda}^{x}(k,p)\right\}R_{\sigma'\tilde\Lambda}(p),\nonumber
\ea
where $\tilde \Lambda=\Max\{|\xi_{\v p}|,|\xi_{\v p+\v Q}|\}$. One can decouple Eq. (\ref{phresp}) into spin- and charge sectors by writing $\chi=\chi^c+\chi^s$ and
\ba
\frac\ud{\ud\Lambda}\,\chi^{c,s}_\Lambda&=&-\frac1{\beta V}\sum_k\frac\ud{\ud\Lambda}\left[C_\Lambda(k)C_\Lambda(k+Q)\right]\,|R^{c,s}_{\tilde\Lambda}(k)|^2,\label{phresponse}\\
\frac\ud{\ud\Lambda}R_\Lambda^{c,s}(k)&=&\frac1{\beta V}\sum_p\frac\ud{\ud\Lambda}\left[C_\Lambda(p)C_\Lambda(p+Q)\right]g_{\tilde\Lambda}^{c,s}(k,p)R_{\tilde\Lambda}^{c,s}(p),\nonumber
\ea
where $R^{c,s}:=R_\up\pm R_\down$, $g^c:=2g^d-g^x$ and $g^s=-g^x$.

\subsection{Relation to parquet diagrams}\label{parquet}

Instead of calculating the effective interaction via the Polchinski equation (\ref{exact}), one could try a naive perturbative expansion (starting from Eq. (\ref{effint})).
$$
g_\Lambda=g_{\Lambda_0}+A_1(\Lambda)\,g^2_{\Lambda_0}+A_2(\Lambda)\,g^3_{\Lambda_0}+\ldots
$$
The problem with this series is that the coefficients $A_n(\Lambda)$ diverge as $\Lambda\to 0$. Their asymptotic behavior is  given by $A_n(\Lambda)\sim l^n(\Lambda)$, where in general $l(\Lambda)=|\log\Lambda|$ except for van Hove filling where $l(\Lambda)=\log^2\Lambda$. 
If the bare interaction $g_{\Lambda_0}$ is small but $g_{\Lambda_0}l(\Lambda)\sim1$, it is a reasonable approximation to sum the whole series, treating every term to leading logarithmic order in $\Lambda$. This amounts to summing the so-called parquet diagrams (see \cite{Nozieres} for a detailed description of this method). 

It is interesting to note that these diagrams are generated by successively integrating Eq. (\ref{1loop}) and expressing the result in terms of the bare interaction $g_{\Lambda_0}$.\footnote{
The structure of Eq. (\ref{1loop}) introduces a constraint on the energies of internal lines. For example, if a parquet diagram is generated by insertion of a bubble $B_1$ into a bubble $B_2$, both single particle energies of $B_1$ have to be bigger than those of $B_2$. The different RG schemes \cite{Drazen,Halboth,Honerkamp} all generate the whole series of parquet diagrams, but the constraints are different.}  We conclude that solving Eq. (\ref{1loop}) to leading order in $\Lambda$ is equivalent to a summation of the parquet diagrams within logarithmic precision.

\section{The RG flow in the asymptotic regime}\label{lead}

The truncation schemes applied so far to Eq. (\ref{exact}) are all of 
perturbative nature and only justified as long as the effective couplings are 
weak. Nevertheless, Eq. (\ref{1loop}) and also its variants are still too 
complicated to be dealt with numerically or analytically. We therefore
 search for a second small parameter, in addition to the coupling strength. 

A natural small parameter is the energy cutoff $\Lambda$ at the final
 stage of the RG flow, where the important contributions come from
 momenta close to the Fermi surface. At this final stage we
 approximate the coupling functions $g_\Lambda(k_1,\ldots,k_4)$ by
 their values on the Fermi surface, i.e., the $2+1$ dimensional
 momenta $k_i$ in $g_\Lambda(k_1,\ldots,k_4)$  are replaced by suitable projections on the Fermi surface ($\xi_{\v k}=k_0=0$). We therefore write $g_\Lambda(\v k_1,\ldots,\v k_4)$ instead of $g_\Lambda(k_1,\ldots,k_4)$, because the frequencies have been set to zero.
 The geometrical constraint of locating all the four momenta $\v k_1,\v k_2,\v k_3$ and $\v k_4=\v k_1+\v k_2-\v k_3$ close to the Fermi surface often reduces the number of variables even further, so that $g_\Lambda$ depends on two or at most three independent (angular) variables instead of nine.   

This restriction to the final stage of the RG flow of the effective couplings implies that we cannot relate the effective theory to the original microscopic Hamiltonian. More importantly, we have to assume 
that no instability occurred in the flow before we have reached the asymptotic 
region.

In the following we identify the leading contributions to Eq. (\ref{1loop}) in
 the limit where these simplifications are justified, i.e., in the asymptotic 
regime $\Lambda\to0$. The external momenta are assumed to lie on the Fermi 
surface.

We first observe that the contributions to Eq.  (\ref{1loop}) are of the
 form 
\be
-\frac1{\beta V}\sum_p\frac{\ud\left[C_\Lambda(p)C_\Lambda(k\mp p)\right]}{\ud\Lambda}g_{
\tilde\Lambda}(\ldots) g_{\tilde\Lambda}(\ldots)\label{form}
\ee
where $k=k_1+k_2$ in diagram PP, $k=k_3-k_1$ in diagram PH1 and $k=k_3-k_2$ in
 diagram PH2. The minus sign appears in the particle-particle (p-p) diagram PP
 and the plus sign in the particle-hole (p-h) diagrams PH1 and PH2.

If we neglect
 for the moment the angular dependence of two coupling functions we are left with an analysis of the 
two bubbles, 
\be
B_{ pp,\,ph}(\Lambda,k)=-\frac1{\beta V}\sum_p\frac{\ud\left[C_\Lambda(p)C_\Lambda(k\mp 
p)\right]}{\ud\Lambda}.\label{bubbles}
\ee
 We consider the thermodynamic limit and zero 
temperature and therefore replace $1/{\beta V}\sum_k$ by
 $\int \frac{d^{2+1}k}{(2\pi)^{2+1}}$ in the calculations. 
Taking explicitly the derivative with respect to $\Lambda$ and integrating over the frequency $p_0$ (for $k_0=0$) we find 
\be
B_{ pp,\,ph}(\Lambda,\v k)= \pm 2\int\!\frac{d^{2}p}{(2\pi)^{2}}\,\delta(|\xi_{\v p}|-\Lambda)\, \frac{\Theta(|\xi_{\v q}|
-\Lambda)\, \Theta(\pm \xi_{\v p}\xi_{\v q})}{\Lambda+|\xi_{\v q}|}, \label{explicit}
\ee
 where $\v q=\v k\mp\v p$.

Within the Wick ordered scheme \cite{Halboth}, the first step function in Eq.
 (\ref{explicit}) is replaced by $\Theta(\Lambda-|\xi_{\v q}|)$, since the 
second propagator in Eq. (\ref{form}) is restricted to be in the low energy part. We have verified that this alternative scheme would not change our final results.

\subsection{Circular Fermi surface}\label{circ}

In order to illustrate the approach we consider the example of very low filling,
 where the single-electron spectrum is approximately parabolic $\xi_{\v p}=\v 
p^2-1$ (all energies are given in units of the Fermi energy and all momenta in 
units of $k_F$). The energy shell consists of two circles with radius $\sqrt{1
\pm\Lambda}$.

In this case, self energy effects are not expected to change the leading order result. On the one hand, the shape of the Fermi surface is fixed by the rotational symmetry of the problem. On the other hand, perturbative corrections to the Fermi velocity $\nabla_{\v p}\Sigma_\Lambda(p)$ and the quasi particle weight $z=(1+i\partial_{p_0}\Sigma_\Lambda(p))^{-1}$ are finite as $\Lambda\to0$.

For general values of $\v k$ one finds that both bubbles $B_{pp,ph}(\Lambda,\v k)$  are proportional to $\log\Lambda$. For the special value $|\v k|=2$ we find $B_{pp}\sim B_{ph}\sim\Lambda^{-1/2}$. The strongest divergence comes from 
the p-p bubble at $\v k=0$, namely
$B_{pp}\sim\Lambda^{-1}$. Correspondingly, the dominant contribution
in the asymptotic regime $\Lambda\to 0$ comes from the diagram PP. It
renormalizes the coupling function $g^{BCS}_\Lambda(\v k,\v
k'):=g_\Lambda(\v k,-\v k,-\v k',\v k')$ which is related to
superconductivity, as was shown in Section \ref{corr}.
The non-locality in $\Lambda$ of Eq. (\ref{1loop}) is absent if only this diagram is taken into account. 

The first sub-leading 
contribution ($\sim\Lambda^{-1/2}$) concerns scattering processes with 
momentum transfer $2k_F$. This contribution is usually neglected. We therefore conclude, in line with the more standard scaling analysis used in previous works \cite{Intro}, that the dominant flow renormalizes only the BCS couplings through the particle-particle diagram.

It is worthwhile to discuss the behavior of the bubbles for small but finite values of $\v k$. For the p-p bubble we find in the limit $|\v k|,\Lambda\ll 1$
\be 
B_{pp}(\Lambda,\v k)=\frac1{2\pi^2\sqrt{|\v k^2-\Lambda^2|}}\left\{\begin{array}{ll}
2\arctan{\sqrt{\frac{\Lambda-|\v k|}{\Lambda+|\v k|}}} & 
\mbox{if }|\v k|<\Lambda,\\
\log{\frac{|\v k|+\sqrt{\v k^2-\Lambda^2}}{\Lambda}} & \mbox{if }|\v k|>\Lambda.
\end{array}\right.
\ee
If we renormalize a coupling with a small total momentum $\v k=\v k_1+\v k_2$ 
the p-p contribution to its flow is independent of $\v k$ as long as 
$\Lambda\gg|\v k|$. In this case we replace $g_\Lambda(k_1,\ldots,k_4)$ by 
$g^{BCS}_\Lambda(\v k_1,\v k_4)$ even if the total momentum is not exactly 
zero. This replacement is no longer justified when $\Lambda$ is of order 
$|\v k|$. The flow then depends strongly on $\v k$ and cannot be controlled. 
Nevertheless, there is no danger because a few renormalization steps later, if 
$\Lambda\ll|\v k|$, the flow is suppressed and the coupling under consideration no longer contributes.

We can use the same type of analysis for the p-h bubble with a small momentum 
transfer, which renormalizes couplings that are close to the forward- or 
exchange scattering $g^f_\Lambda(\v k,\v k'):=Xg^e_\Lambda(\v k,\v k'):=g_\Lambda(
\v k,\v k',\v k',\v k)$. We obtain 
\be 
B_{ph}(\Lambda,\v k)=\left\{\begin{array}{ll}
0 & \mbox{if }|\v k|<\Lambda,\\
-\frac1{2\pi^2|\v k|}\log{\frac{|\v k|+\sqrt{\v k^2-\Lambda^2}}{\Lambda}} & 
\mbox{if }|\v k|>\Lambda.
\end{array}\right.
\ee
It gives a big $\v k$-dependent contribution if $\v k$ is of order 
$\Lambda$. But this flow is again suppressed if $\Lambda$ is further reduced. 

The results presented above for the isotropic case should remain valid as long as the Fermi 
surface is both far away from van Hove singularities and not nested. Even the 
presence of Umklapp scattering alone does not change the result. If the Fermi 
surface is big enough, Umklapp processes open up new scattering processes, but 
their contribution to the flow is not of leading order \cite{Rajaraman}. 
They can nevertheless influence the flow in an important way before we enter 
the asymptotic regime \cite{Honerkamp}. But at 
this early state of the RG flow, approximations to Eq. (\ref{1loop}) cannot be 
controlled by the small parameter $\Lambda$.

\subsection{Generic Fermi surface at van Hove filling}\label{vHove}

\begin{figure}
        \centerline{\includegraphics[width=4cm]{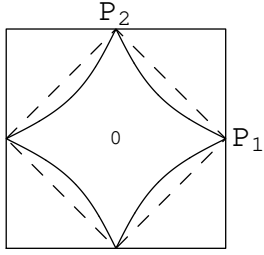}}
\caption{The Fermi surface at van Hove filling: $t'=0.3$ (solid line) and $t'=0$ (dashed line).}
\label{ttprimeFS}
\end{figure}

We now consider the case where the Fermi surface passes through a van Hove singularity.
The generic situation occurs for a band structure with saddle points at $\v P_1=(\pm\pi,0)$ and $\v P_2=(0,\pm\pi)$. To be specific, we consider the dispersion relation of a generalized tight-binding model: $\xi_{\v k}=-2(\cos k_x+\cos k_y)+4t'(\cos k_x\cos k_y+1)$. The unit of energy is given  by the hopping amplitude between nearest neighbors 
and the unit of momenta is the inverse of the lattice constant. A finite electron 
hopping $0<t'<1/2$ between next nearest neighbors has been included 
and the chemical potential is fine-tuned such that the Fermi surface contains 
the saddle points (see Fig. \ref{ttprimeFS}).

We identify again the dominant terms of order $\Lambda^{-1}$ in the RG equation (\ref{1loop}) by analyzing the bubbles  $B_{pp}(\Lambda,\v k)$ and $B_{ph}(\Lambda,\v k)$. Dominant terms of this order appear only if $\v k$ is close to $\v 0$ or close to $(\pi,\pi)$.

We first study the behavior of the bubbles at small $|\v k|$.  
It is instructive to focus on the contributions of a small patch surrounding a saddle point, say the region  
$$P_1=\left\{\v p\ ;\sqrt{1-2t'}\,|p_x-\pi|+\sqrt{1+2t'}\,|p_y|\leq 2\rho\right\}.$$ 
The parameter $\rho$ is small enough so that 
$\xi_{\v p}$ can be replaced by its limiting quadratic form close to $\v P_1$.
 We compute the restricted bubbles $B_{pp}^{P_1}(\Lambda,\v k)$ and $B_{ph}^{P_1}(\Lambda,\v k)$, which are defined by Eq. (\ref{explicit}), but restricting the summation to $\v p\in P_1$. The values of  $B_{pp}^{P_1}(\Lambda,\v k)$ and $B_{ph}^{P_1}(\Lambda,\v k)$ depend sensitively on $\xi_{\v P_1+\v k}$. Both bubbles are negligible if $|\xi_{\v P_1+\v k}|\gg \Lambda$, but for $|\xi_{\v P_1+\v k}|\ll \Lambda$ (and $\Lambda,\v k^2\ll\rho^2$) we get
\be
B_{pp}^{P_1}(\Lambda,\v k)=\frac1{(2\pi)^2\sqrt{1-4t'^2}\,\Lambda}\ \log\frac{4\Lambda\rho^2}{(\Lambda+\rho\, k_+)(\Lambda+\rho\, k_-)}\label{ppvH}
\ee
and
\be
B_{ph}^{P_1}(\Lambda,\v k)=\frac{-1}{(2\pi)^2\sqrt{1-4t'^2}\,\Lambda}\left[\theta(\rho\,k_+-\Lambda)\frac{\rho\,k_+-\Lambda}{\rho\,k_+}+\theta(\rho\,k_--\Lambda)\frac{\rho\,k_--\Lambda}{\rho\,k_-}\right],\label{phvH}
\ee
where $k_\pm:=|\sqrt{1-2t'}\,k_x\pm\sqrt{1+2t'}\,k_y|$. Note that $k_+\,k_-=|\xi_{\v P_1+\v k}|$.

There are several remarkable differences with respect to the circular
Fermi surface. First we observe that for very small $\v k$, such that $\rho\,k_\pm\ll\Lambda$, we have $B_{pp}^{P_1}(\Lambda,\v k)\sim\Lambda^{-1}\log\frac{4\rho^2}{\Lambda}$, i.e. the p-p bubble has a logarithm in addition to the $\Lambda^{-1}$ behavior. This anomaly, due to the diverging density of states at the Fermi level, concerns only the renormalization of the BCS couplings. 

More striking is the behavior in the regime $k_+k_-\ll\Lambda\ll\rho\,k_\pm$. The p-h bubble, zero for 
sufficiently small $|\v k|$, gives a non negligible contribution there. Thus 
for a given small momentum transfer $\v k$, the p-h contributions [diagram PH1 
or PH2 of Fig. \ref{diags}] are dominant over many RG iterations, in contrast 
to the case of Section \ref{circ} without van Hove singularities. The p-p diagram 
is very sensitive with respect to both the size and the direction of the total
 momentum $\v k$ in this regime, i.e. over a 
sizeable range of energy scales. We conclude that in a consistent treatment of the RG equations to order 
$\Lambda^{-1}$, the renormalization of the effective coupling function
$g_\Lambda(\v k_1,\dots,\v k_4)$ depends on the exact values of its arguments. For example the abovementioned projection of the momenta onto the Fermi surface cannot be justified in the vicinity of a van Hove point.

One way to reduce the complexity of the problem is to keep only the terms of 
 order $\Lambda^{-1}\log(1/\Lambda)$ and to neglect all other contributions.
This leading logarithmic contribution occurs only in the 
p-p bubble for $|\v k|\ll\sqrt\Lambda$. As a consequence only the BCS 
coupling function $g^{BCS}_\Lambda(\v k,\v k')$ is renormalized within this approximation.
The flow of  $g^{BCS}_\Lambda(\v k,\v k')$ is similar to the one explained in detail in the next section, but without
 competing charge and spin instabilities. For $\v k$ close to $(\pi,\pi)$ and finite values of $t'$ the p-p and p-h bubbles behave like $\Lambda^{-1}$ but not like  $\Lambda^{-1}\log\Lambda$.

For repulsive interactions, where the BCS flow is towards weak coupling, the 
approximation is clearly not sufficient and a consistent treatment up to
 order $\Lambda^{-1}$ is required. Solutions for this case proposed earlier have neglected  the sensitive dependence of $g(\v k_1,\ldots,\v k_4)$ on its arguments \cite{Lederer,Dzyalo2,Furukawa}. 

In contrast to the case of the circular Fermi surface, self-energy corrections are not negligible in the present situation. In fact, they are expected to change the shape of the Fermi surface.

For the corrections to the single particle dispersion and the quasi particle weight, one finds in second order perturbation theory (see for example Eq. 7 of \cite{Dzyalo2})
\ba
\partial_{p_0}\Sigma_\Lambda(p)&\sim&g^2_{\Lambda_0}\log^2\Lambda\label{SEcorr}\\
\nabla_{\v p}\Sigma_\Lambda(p)&\sim&g^2_{\Lambda_0}\log^2\Lambda\,\cdot\,\nabla_{\v p}\xi_{\v p}\nonumber
\ea
These corrections are negligible within our approximation, as long as $g_{\Lambda_0}\log^2\Lambda\sim1$. An attractive coupling diverges at this scale. In the case of repulsive interactions however the RG flow can be followed to smaller energies such that $g_{\Lambda_0}\log\Lambda\sim1$. At these energy scales, self energy corrections have to be taken seriously. This would require to include subleading terms, a task that cannot be fulfilled consistently without going beyond the one-loop approximation.

Due to the additional logarithm in the p-p bubble the theory is not
renormalizable in the usual sense of field theory, i. e. it is
not possible to send the bare momentum cutoff to infinity while
keeping some physical correlation functions finite (even after the
introduction of counter-terms in the microscopic Hamiltonian and of
wave-function renormalization). Gonzalez, Guinea and 
Vozmediano \cite{Madrid-vH} proposed a field theoretical RG scheme where the coupling
 constants for forward and exchange scattering are renormalized only 
by the p-h diagrams. In this approach the p-p channel is treated
separately in connection with a renormalized chemical potential. It is
however not clear from our Wilsonian approach that the RG equations 
will not mix p-p and p-h diagrams, if all contributions 
$\sim\Lambda^{-1}$ are taken into account.

\section{The square Fermi surface}\label{squareFS}

\begin{figure} 
        \centerline{\includegraphics[width=10cm]{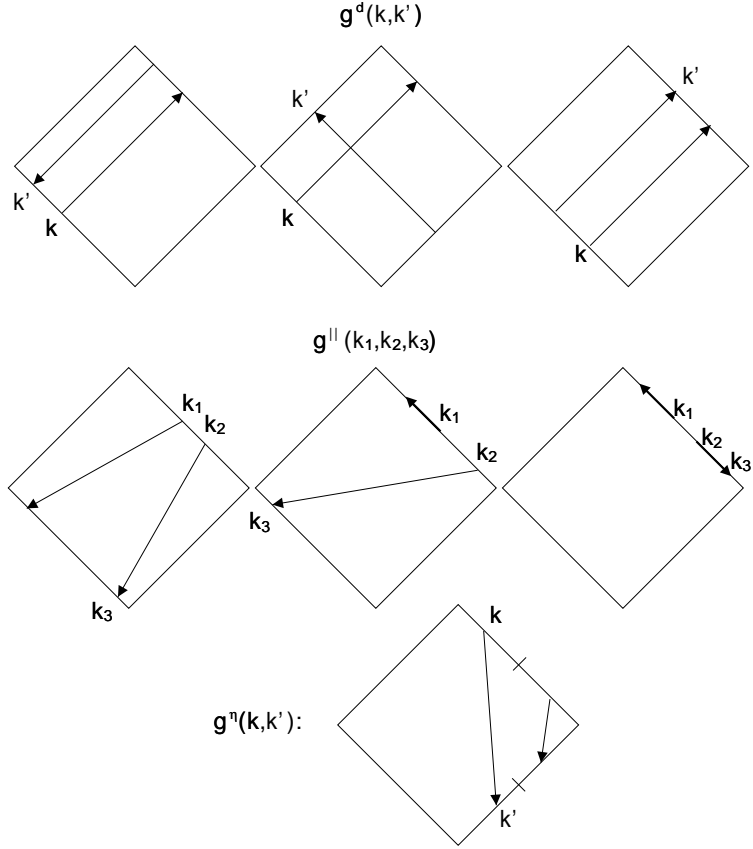}}
\caption{Examples of processes from the different interaction channels}
\label{channels}
\end{figure}

In this and in the following Sections we consider the nearest-neighbor tight-binding model ($t'=0$) at half filling where 
the Fermi surface is a square (see Fig. \ref{ttprimeFS}).
We first identify the possible scattering processes which connect four momenta on the Fermi surface and satisfy momentum conservation modulo a reciprocal lattice vector. To simplify notation we omit the subscript $\Lambda$ in $g_\Lambda$. In addition to the usual forward, exchange and BCS scattering, there are the
two kinds of processes related to direct or exchanged transfer of the nesting 
vector $\v Q=(\pi,\pi)$: $g^d(\v k,\v k'):=Xg^x(\v k,\v k')=g(\v k,\v k'+
\v Q,\v k',\v k+\v Q)$. Furthermore if three points $\v k_1,\v k_2,\v k_3$ are 
chosen freely on two parallel sides of the square, the resulting $\v k_4=\v
 k_1+\v k_2-\v k_3$ lies automatically on the Fermi surface, giving rise to
 a three-parameter family $g^\parallel(\v k_1,\v k_2,\v k_3)$. Finally, there
 is scattering of pairs with a total momentum $\v Q$, described by 
$g^\eta(\v k,\v k'):=g(\v k,\v Q-\v k,\v Q-\v k',\v k')$. Some examples are shown in Fig. \ref{channels}.  

This list of the possible low energy processes is complete but the
classification into $g^f$, $g^e$, $g^{BCS}$, $g^d$, $g^x$, $g^\eta$
and $g^\parallel$ is not unique. For example   if $\v k$ and $\v k'$
belong to the same pair of parallel sides of the square, the
two-parameter families $g^f$, $g^e$, $g^{BCS}$, $g^d$, $g^x$ and
$g^\eta$ belong to the larger three parameter family $g^\parallel$. Furthermore, two two-parameter families intersect in a one-parameter family as $g^{BCS}(\v k,\v k+\v Q)=g^{d}(\v k,\v Q-\v k)$,  $g^{BCS}(\v k,\v Q-\v k)=g^{x}(\v k,\v Q-\v k)$, etc. Finally three two-parameter families can intersect in scatterings between the two saddle points as $g^{BCS}(\v P_1,\v P_2)=g^d(\v P_1,\v P_2)=g^x(\v P_1,\v P_2)$. 

Due to  exact nesting, $\xi_{\v p+\v Q}=-\xi_{\v p}$, the non interacting model has particle-hole symmetry. We further assume that the interaction also respects this symmetry (see Section \ref{symmsection}). The self energy then satisfies the exact relation $\Sigma_\Lambda(p_0,\v p)=-\Sigma_\Lambda(-p_0,\v p+\v Q)$, and therefore vanishes on the Fermi surface ($p_0=\xi_{\v p}=0$). This means that the Fermi surface is not modified by self-energy effects.

For the corrections to the single particle dispersion and the quasi particle weight, one can derive similar relations as in Eq. (\ref{SEcorr}), namely
\ba
\partial_{p_0}\Sigma_\Lambda(p)&<&\mbox{const}\,\cdot\,g^2_{\Lambda_0}\log^3\Lambda\\
\nabla_{\v p}\Sigma_\Lambda(p)&<&\mbox{const}\,\cdot\,g^2_{\Lambda_0}\log^3\Lambda\,\cdot\,\nabla_{\v p}\xi_{\v p}.\nonumber
\ea
 These corrections are negligible within our approximation, where $g_{\Lambda_0}\log^2\Lambda\sim1$ in the spirit of the parquet approximation (see Section \ref{parquet}).

We now investigate the one-loop corrections to the coupling constants by an analysis of the bubbles (\ref{explicit}). As a consequence of the particle-hole symmetry, the p-p and p-h bubbles are related by 
\be 
B_{pp}(\Lambda,\v k)=-B_{ph}(\Lambda,\v k+\v Q).\label{nesting}
\ee

\begin{figure} 
        \centerline{\includegraphics[width=10cm]{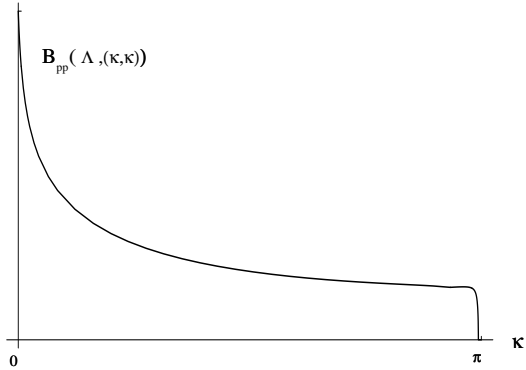}}
\caption{Plot of the p-p bubble for $\v k=(\kappa,\kappa)$, parallel to the square Fermi surface and $\Lambda=0.01$.}
\label{Bpp}
\end{figure}

Both the p-p and p-h bubbles are of order  $\Lambda^{-1}$  whenever $\v k=n(\pi,\pi)+\kappa(1,\pm1)$ ($n\in\Z,\kappa\in\R$). It follows that the three parameter coupling function $g^\parallel $ is renormalized by 
contributions of order $\Lambda^{-1}$ from every diagram PP, PH1 and PH2 
in Eq. (\ref{1loop}). The reason is that the Fermi surface
 consists of straight lines. Namely if $\v k$ is parallel to  such a line, 
$\xi_{\v p}$ and $\xi_{\v p+\v k}$ are both small for many values of $\v p$,
 which gives rise to a big value of the integral in Eq. (\ref{explicit}). In Fig. \ref{Bpp} we show a plot of $B_{pp}(\Lambda,\v k)$ for $\v k=(\kappa,\kappa)$. For $\Lambda\ll\kappa\ll1$, $B_{pp}(\Lambda,\v k)$ is well approximated by $(2\pi)^{-2}\Lambda^{-1}\log\frac{4}\kappa$, whereas for $\kappa\ll\Lambda$ it approaches its maximal value  $B_{pp}(\Lambda,\v 0)=(2\pi)^{-2}\Lambda^{-1}\log\frac{16}\Lambda$.

Due to this big variety of  contributions of order  $\Lambda^{-1}$,
the flow to that order appears to be too complicated for analytical treatment. Numerical treatments have been 
presented by Zanchi and Schulz \cite{Drazen} and more recently by Halboth
 and Metzner \cite{Halboth}. A parquet solution for a flat Fermi surface 
 has been given by Zheleznyak, Yakovenko and Dzyaloshinskii \cite{ZYD}.

By restricting ourselves to the logarithmically dominant
 terms of order $\Lambda^{-1}\log\Lambda$ we can go a long way
 using an analytical approach, as will be shown now. We consider the two-parameter coupling functions $g^{BCS}$, $g^x$ and $g^d$. $g^{BCS}$ gets a dominant contribution ($\sim B_{pp}(\Lambda, 0)$) from the diagram PP,  $g^x$ from PH1 and $g^d$ from PH2. For example we have 
$$
\frac\ud{\ud\Lambda}g^{BCS}(\v k,\v k')\ =\ \mbox{PP}+\mbox{PH1}+\mbox{PH2},
$$
where $\mbox{PP}\sim B_{pp}(\Lambda, 0)\sim\Lambda^{-1}\log(1/\Lambda)$,  $\mbox{PH1}\sim B_{ph}(\Lambda, \v k+\v k')$ and  $\mbox{PH2}\sim B_{ph}(\Lambda, \v k-\v k')$. We may neglect PH1 and PH2 for small values of the cutoff, except when $\v k\pm\v k'\approx\v Q$. Similarly, for  generic $\v k$ and $\v k'$, $g^x(\v k,\v k')$
 is renormalized only by the diagram PH1 and $g^d(\v k,\v k')$ only 
by PH2. The flow equations (\ref{1loop}) are again local in $\Lambda$ and 
they decouple into three independent identical equations for the functions 
$g^{BCS}$, $g^s=-g^x$ and $g^c=2g^d-g^x$. They read 
\be
2\Lambda\frac\ud{\ud\Lambda}g^{\diamond}(\v k,\v k')=\frac1V\sum_{\v p}
\delta(|\xi_{\v p}|-\Lambda)\,g^{\diamond}(\v k,\v p) g^{\diamond}
(\v p,\v k')\label{rg},
\ee
where $\diamond$ stands for $BCS$, $s$ or $c$. This is
 evident for $g^{BCS}$ and $g^s$ since only one diagram is
 involved in their renormalization. The PH2 contribution to $g^d$ on the other hand consists of
three terms (the three PH2 diagrams) involving $g^d$ and $g^x$. We see with amazement that the
 special combination $g^c=2g^d-g^x$ satisfies the same closed and simple equation as $g^{BCS}$
 and $g^s$. It was shown in Section \ref{corr} that $g^{BCS}$ is related to pairing, $g^s$ to spin instabilities and $g^c$ to charge instabilities.

\begin{figure} 
        \centerline{\includegraphics[width=5cm]{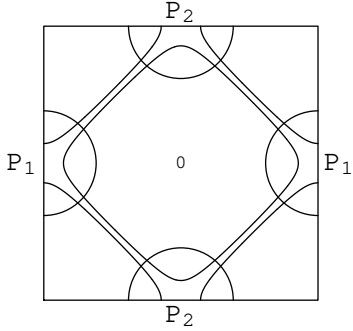}}
\caption{The Brillouin zone with the energy shell $|\xi_{\v p}|=\Lambda$. It is
 separated into two patches $P_1$ and $P_2$ and the remaining part.}\label{bz}
\end{figure}

In the integration over $\v p$ in Eq. (\ref{rg}) a large contribution comes from a small neighborhood of the saddle-points, due to the diverging density of states. To identify the logarithmically diverging contribution to Eq. (\ref{rg}) we consider patches $P_1=\left\{\v p\,;\,|p_x-\pi|+|p_y|<2\rho\right\}$ and $P_2=\left\{\v p\,;\,|p_x|+|p_y-\pi|<2\rho\right\}$ of a size
 $\rho\ll\pi/2$ around the two van Hove points and separate the integral 
$\sum_{\v p}$ into  $\sum_{\v p\in P_1}+\sum_{\v p\in P_2}+\sum_{\v p\in 
B.Z.-P_1-P_2}$, where $B.Z.$ is the whole Brillouin zone (see Fig. \ref{bz}).
 We compute the weight of the patch and of the remaining part of the Brillouin 
zone, assuming that  $g^\diamond(\v k,\v p)$ is of the same order of magnitude 
for every value of $\v p$. Comparing the values
\be 
\frac1V\sum_{\v p\in P_1}\delta(|\xi_{\v p}|-\Lambda)=\frac1{2\pi^2} \log(\frac{4\rho^2}{\Lambda}),\label{p1}
\ee
\be 
\frac1V\sum_{\v p\in B.Z.-P_1-P_2}\!\delta(|\xi_{\v p}|-\Lambda)=\frac1{\pi^2}
\log(\frac4{\rho^2}),\label{p2}
\ee
 we conclude that the patch contribution dominates the remaining part 
if $\Lambda\ll\rho^4$. Under the hypothesis that the functions $g^\diamond(\v k,\v k')$ are slowly varying, it is then consistent to replace Eq. (\ref{rg}) by
\be
\frac\ud{{\ud} l}g^\diamond(\v k,\v k')=-\sum_{i=1,2}\,g^\diamond_i(\v k) 
g^\diamond_i(\v k'),\label{prg}
\ee
where we have set $g^\diamond(\v k,\v p)\approx g^\diamond_i(\v k)$ for 
$\v p\in P_i$ and $\;l=\frac1{8\pi^2}\log^2(\frac{4\rho^2}{\Lambda})$ as the new RG parameter.
Correspondingly, setting $g^\diamond_i(\v k')\approx g^\diamond_{ij}$
 for  $\v k'\in P_j$ in (\ref{prg}) leads to 
\be
\frac\ud{{\ud}l}g^\diamond_j(\v k)=-\sum_{i=1,2} \,g^\diamond_i(\v k) \,
g^\diamond_{ij}.\label{prg2}
\ee

We are now left with the problem of renormalizing the couplings
$g^\diamond_{ij}$, which describe scattering of particles near the
saddle points. Eq. (\ref{rg}) cannot be used here because it is not
valid for $\v k\pm\v k'\approx\v Q$, i.e. if
$\v k\in P_1$ and $\v k'\in P_2$. The RG flow of the couplings
$g^\diamond_{ij}$ will be discussed in detail in Section \ref{2patch}.

Once the evolution of the $g^\diamond_{ij}$ as a function of $l$ is 
known, one can integrate Eqs. (\ref{prg2}) and (\ref{prg}). 
Introducing $g^\diamond_\pm(\v k)=g^\diamond_1(\v k)\pm g^\diamond_2(\v k)$
 we get
\be
g^\diamond_\pm(\v k) =\left.g^\diamond_\pm(\v k)\right|_{l=l_0}\,\cdot\,
\exp\left[-\int_{l_0}^l{\ud} l\,\left(g^\diamond_{11}\pm 
g^\diamond_{12}\right)\right]\label{pprg2}
\ee
and
\be
g^\diamond(\v k,\v k')=\left.g^\diamond(\v k,\v k')\right|_{l=l_0}-
\frac12\int_{l_0}^l{\ud} l\,\left[g^\diamond_+(\v k)g^\diamond_+(\v k')
+g^\diamond_-(\v k)g^\diamond_-(\v k')\right].\label{pprg}
\ee

We will see that the RG equations of $g^\diamond_{ij}$ as a function of $l$ yield diverging solutions at a finite value $l=l_c$. Near this critical value they behave asymptotically like 
\be 
g^\diamond_{ij}(l)\approx\frac{\tilde g^\diamond_{ij}}{l_c-l}+O(l_c-l)^\alpha,\label{divergentg}
\ee
where the constant $\tilde g^\diamond_{ij}$ can be determined from the RG equation and $\alpha>-1$. Eqs. (\ref{pprg2}) and (\ref{pprg}) then give
\be
g^\diamond_\pm(\v k)\approx A_\pm\cdot\left.g^\diamond_\pm(\v k)\right|_{l=l_0}\,\cdot\,
\left[(l_c-l)^{\tilde g^\diamond_{\pm}}+O(l_c-l)^{\tilde g^\diamond_{\pm}+\alpha+1}\right]
\ee
and
\be
g^\diamond(\v k,\v k')\approx\left.g^\diamond(\v k,\v k')\right|_{l=l_0}+\sum_{\nu=\pm}B_\nu\left.g^\diamond_\nu
(\v k)g^\diamond_\nu(\v k')\right|_{l=l_0}
\left[(l_c-l)^{2\tilde g^\diamond_{\nu}+1}+O(l_c-l)^{\Min{\{0,2\tilde g^\diamond_{\nu}+\alpha+2\}}}\right],\label{gkksol}
\ee
where $\tilde g^\diamond_{\pm}:=\tilde g^\diamond_{11}\pm\tilde g^\diamond_{12}$ and $A_\pm$, $B_\pm$ are positive constants.
The coupling function 
$g^\diamond(\v k,\v k')$ is diverging if $\tilde g^\diamond_{+}$ or $\tilde g^\diamond_{-}\leq-1/2$. 

The functions $g^\diamond_+(\v k)$ have $s$-wave symmetry, i.e. they
respect all the point symmetries of the square lattice. On the other hand we see that
$g^\diamond_-(\v k)$ is of the $d_{x^2-y^2}$-wave type 
[$g^\diamond_-(k_x,k_y)=g^\diamond_-(k_x,-k_y)=-g^\diamond_-(k_y,k_x)$].
 Thus the diverging part of the coupling 
function has $s$ or $d_{x^2-y^2}$-wave symmetry.

In the preceding calculation we have distinguished strictly between points far from the saddle points and those close to the saddle points. The scale which distinguishes between ``far'' and ``close'' is the patch size $\rho$, which was introduced by hand. 

The behavior of the overall coupling function $g^\diamond(\v k,\v k')$
near the critical point depends on the smallest (i.e. the most
negative) value of the constants $\tilde g^\diamond_\pm$. In Section
\ref{our} we will show that 
$\Min\{\tilde g^\diamond_\pm\}=-1$. In this situation the function $g^\diamond(\v k,\v
k')$ diverges everywhere with the same power $(l-l_c)^{-1}$ and  
$g^\diamond(\v k,\v k')$ remains a smooth function upon
renormalization even if $\v k$ or $\v k'$ (or both) approach the
saddle points. 
This justifies a posteriori the estimation of the
diagrams (\ref{form}) by the bubbles (\ref{bubbles}) as well as the 
approximations of Eqs. (\ref{prg}) and (\ref{prg2}), namely that $g^\diamond(\v k,\v k')$ approaches continuously $g^\diamond_i(\v k)$ as $\v k'\to \v P_i$ and $g^\diamond_i(\v k)$ approaches $g^\diamond_{ij}$ as  $\v k \to \v P_j$.
If instead we would find $0>\Min\{\tilde g^\diamond_\pm\}>-1$, the scattering of particles
near the van Hove points would diverge more rapidly than that of particles
at the remaining Fermi surface. This would mean that different regions
in the Brillouin zone behave differently. The region around the
saddle points would become strongly interacting while the remaining
Fermi surface would remain weakly interacting.

In order to calculate the susceptibilities we consider Eqs.  (\ref{ppresponse}) and (\ref{phresponse}). The non-locality in the variable $\Lambda$ of Eq. (\ref{phresponse}) disappears because of exact nesting ($\xi_{\v p+\v Q}=-\xi_{\v p}$).
In the low energy  regime $\Lambda\to0$, we assume that the frequency
dependence of $R$ is not important and replace $R(k)$ by $R(\v k)$. We can then explicitly perform the frequency integral to obtain identical equations for the charge, spin and pairing susceptibilities
\ba
2\Lambda\frac\ud{\ud\Lambda}\,\chi^\diamond&=&-\frac1{V}\sum_{\v k}\delta(|\xi_{\v k}|-\Lambda)\,|R^{\diamond}(\v k)|^2,\nonumber\\
2\Lambda\frac\ud{\ud\Lambda}R^{\diamond}(\v k)&=&\frac1{V}\sum_{\v p}\delta(|\xi_{\v p}|-\Lambda)\,g^\diamond(\v k,\v p)R^\diamond(\v p),\label{response}
\ea
where $\diamond=c,s,BCS$. To simplify the notation we have omitted the index $\Lambda$ in the symbols $\chi$, $R$ and $g$.

We treat Eq. (\ref{response}) in the same way as Eq. (\ref{rg}), i.e. we take the leading contribution from the patches $P_1,P_2$ around the two saddle points and assume $R^\diamond(\v p)\approx R^\diamond_i$ for $\v p\in P_i$. We get
\ba
\frac\ud{\ud l}\,\chi^\diamond_\pm&=&\frac12|R^{\diamond}_\pm|^2,\nonumber\\
\frac\ud{\ud l}R^{\diamond}_\pm&=&-g^\diamond_{\pm} R^\diamond_\pm,\label{presponse}
\ea
where $R^\diamond_\pm:=R^\diamond_1\pm R^\diamond_2$ and $\chi^\diamond=:\chi^\diamond_++\chi^\diamond_-$. 
If $g^\diamond_\pm$ is diverging asymptotically like $\tilde
g^\diamond_\pm/(l_c-l)$ and if $\tilde
g^\diamond_\pm<-1/2$, the corresponding susceptibility diverges with a critical exponent $2\tilde g^\diamond_\pm+1$:
\be
\chi^\diamond_\pm\sim(l_c-l)^{2\tilde g^\diamond_\pm+1}\sim(\Lambda-\Lambda_c)^{2\tilde g^\diamond_\pm+1}.\label{crit}
\ee
On the other hand $\tilde g^\diamond_\pm\geq-1/2$ leads to a finite
value of the susceptibility.

We thus naturally identify six possible instabilities. In each of the
charge, spin or pairing sectors the form factor can be either even or
odd under the exchange of the two van Hove points (i.e. $R_1=R_2$ or
$R_1=-R_2$). In the pairing sector $\chi^{BCS}_+$ clearly corresponds
to $s$-wave superconductivity (sSC)  and $\chi^{BCS}_-$ to
$d_{x^2-y^2}$-wave  superconductivity (dSC). The charge and spin density waves are related to the susceptibilities 
$\chi_+^c$ and $\chi_+^s$, respectively. The two remaining
susceptibilities $\chi^{c,s}_-$ correspond to a form factor with
$d_{x^2-y^2}$-wave symmetry in the charge and spin sectors. They
describe the tendency towards the formation of charge or spin flux
phases (see Section \ref{corr}).

These six instabilities of a system with two van Hove 
singularities have been discussed long ago by H. J. Schulz \cite{Schulz2}. 
Here we have shown that they appear naturally in the very weak coupling 
limit of Wilson's renormalization group.

\section{Flow of the couplings between saddle points.}\label{2patch}

We now return to the renormalization of coupling constants for scattering processes both within and between saddle point patches. We consider the nearest-neighbor tight-binding model ($t'=0$) as in Section \ref{squareFS}.
The most simple assumption is to treat the saddle points in close analogy to the Fermi points of a one-dimensional system where there are just four types of scattering processes, one restricted to the region of a single Fermi point, the other three involving both right and left movers (forward, backward and Umklapp scattering).
This one-dimensional scenario with four coupling constants $g_1,\ldots,g_4$ implicitly assumes that the more detailed wave vector dependence in this region is irrelevant. While this appears to be true for one-dimensional Fermi systems, where going away from a Fermi point means leaving the Fermi surface, we will argue that in the present case the functional dependence of the couplings is relevant in the neighborhood of the saddle points.

\subsection{``One-dimensional'' solution}\label{historical}

In an early contribution to this subject H. J. Schulz \cite{Schulz1} assumed that the coupling function $g(\v k_1,\ldots,\v k_4)$ takes only four different values
according to how the momenta $\v k_1,\ldots,\v k_4$ are distributed over the two patches, namely 
\be
g(\v k_1,\ldots,\v k_4)\equiv\left\{\begin{array}{ll} 
g_1\;;\v k_1,\v k_3\in P_1\mbox{ and }\v k_2,\v k_4\in P_2\\ 
g_2\;;\v k_1,\v k_4\in P_1\mbox{ and }\v k_2,\v k_3\in P_2\\
g_3\;;\v k_1,\v k_2\in P_1\mbox{ and }\v k_3,\v k_4\in P_2\\ 
g_4\;;\v k_1,\ldots,\v k_4\in P_1
\end{array}\right.,\label{g14}
\ee
Note that we have interchanged $g_2$ and $g_4$ as compared to Ref. \cite{Schulz1}. The parameters $g^\diamond_\pm=g^\diamond_{11}\pm g^\diamond_{12}$, which control the various instabilities ($\diamond=s,c,BCS$), are readily expressed in terms of the coupling constants $g_1,\ldots,g_4$,
\ba
SDW/SF:&\ g_\pm^s&=-g_2\mp g_3\nonumber\\
CDW/CF:&\ g_\pm^c&=2g_1-g_2\pm g_3\nonumber\\
sSC/dSC:&\ g_\pm^{BCS}&=g_4\pm g_3.\label{instab}
\ea

$g_1$ has to be renormalized by the diagrams PH2, because the direct momentum transfer $\v k_3-\v k_2$ is close to the nesting vector $\v Q$, but the other contributions coming from PH1 and PP are negligible. Similarly $g_2$ is renormalized only by PH1 and $g_4$ by PP. The remaining coupling $g_3$ on the other hand gets leading contributions from all three channels, PP, PH1 and PH2.  

The RG equation is obtained by locating the external momenta $\v k_1,\ldots,\v k_4$ in Eq. ({\ref{1loop}) exactly at the saddle points $\v P_1$ and $\v P_2$ and restricting the sum over $\v p$ to the two patches $P_1,P_2$. The result is 
\ba
\frac\ud{\ud l}g_1&=&2g_1(g_2-g_1)\nonumber\\
\frac\ud{\ud l}g_2&=&g_2^2+g_3^2\nonumber\\
\frac\ud{\ud l}g_3&=&2g_3(2g_2-g_1-g_4)\nonumber\\
\frac\ud{\ud l}g_4&=&-g_3^2-g_4^2.\label{differential}
\ea    

For most initial conditions the numerical solution of these equations diverge asymptotically like in Eq. (\ref{divergentg}) with coefficients $\tilde g_i$ satisfying
\ba
\tilde g_1&=&2\tilde g_1(\tilde g_2-\tilde g_1)\nonumber\\
\tilde g_2&=&\tilde g_2^2+\tilde g_3^2\nonumber\\
\tilde g_3&=&2\tilde g_3(2\tilde g_2-\tilde g_1-\tilde g_4)\nonumber\\
\tilde g_4&=&-\tilde g_3^2-\tilde g_4^2.\label{algebra}
\ea   
Eq. (\ref{algebra}) has many solutions, but the ones which are relevant for the divergences of Eq. (\ref{differential}) are $\tilde g_1=0$, $\tilde g_2=-\tilde g_4=1/6$ and $\tilde g_3=\pm\sqrt5/6$, depending on whether the initial value of $g_3$ is positive or negative (note that $g_3$ cannot change its sign). The special feature of these two solutions is that in view of Eqs. (\ref{crit}) and (\ref{instab}) three out of the six dominant susceptibilities are diverging with the same critical exponent. Namely
\ba
\chi_{SDW}\sim\chi_{dSC}\sim\chi_{CF}&\sim(\Lambda-\Lambda_c)^{-\gamma}\quad\mbox{ if }\quad g_3>0,\nonumber\\
\chi_{CDW}\sim\chi_{sSC}\sim\chi_{SF}&\sim(\Lambda-\Lambda_c)^{-\gamma}\quad\mbox{ if }\quad g_3<0,\label{g3discr}
\ea
where $\gamma=(\sqrt5-2)/3\approx0.08$. The divergence of the susceptibilities is thus weak compared to the mean field behavior $\chi\sim(T-T_c)^{-1}$.

However, for some initial conditions the solutions of Eq. (\ref{differential}) are not diverging but flow towards the trivial fixed point $g_1=g_2=g_3=g_4=0$. This means that the RG flow of the couplings between the two saddle point patches does not develop an instability. In this case the restriction to the saddle-point patches is clearly insufficient and the low energy behavior is controlled by the remaining part of the Fermi surface.

\subsection{Towards functional renormalization}\label{our}

The hypothesis of constant couplings $g_1,\ldots,g_4$ neglects the fact that in reality all these parameters are functions of incoming and outgoing momenta $\v k_1,\ldots,\v k_4$, three of which can move freely on the Fermi surface within the saddle point patches. Unfortunately, a true functional renormalization is presently beyond the reach of an analytical approach. Therefore we mimic the true momentum dependence by introducing, in addition to the constants $g_1,\ldots,g_4$ representing general values of the momenta (in the patches), other couplings corresponding to specific combinations of momenta. Thus we allow $g_3(\v k_1,\ldots,\v k_4)$ to take four different values:
\be 
g_3(\v k_1,\ldots,\v k_4)\approx\left\{\begin{array}{ll}
{ g_3^{BCS}}&\mbox{if }\ \v k_1+\v k_2= \v 0\\
{ g_3^x}&\mbox{if }\ \v k_3-\v k_1= \v Q\\
{ g_3^d}&\mbox{if }\ \v k_3-\v k_2= \v Q\\
g_3&\mbox{if }\ |\v k_3-\v k_2- \v Q|,\ |\v k_3-\v k_1 -\v Q|,\ |\v k_1+\v k_2|>O(\sqrt{\Lambda})
\end{array}\right.\label{specialcouplings}
\ee
Similarly, $g_1(\v k_1,\ldots,\v k_4)=$ takes the values  {$g_1^d$} or
 $g_1$,  $g_2(\v k_1,\ldots,\v k_4)={ g_2^x}$ or $g_2$ and  $g_4(\v
 k_1,\ldots,\v k_4)={ g_4^{BCS}}$ or $g_4$. We thus separate the
 couplings $g^{BCS}_3,g^{BCS}_4$ with zero total momentum, $g^d_1$,
 $g^d_3$ with a direct momentum transfer equal to $\v Q$ and
 $g^x_2,g^x_3$ with an exchanged momentum transfer of $\v Q$ from the
 general ones ($g_1,\ldots,g_4$), where none of these special
 relations among the in- and outgoing momenta applies\footnote{The special couplings are related to the couplings $g^\diamond_{ij}$
 introduced in Section \ref{squareFS} by $g^c_{11}=2g^d_1-g^x_2$,
 $g^s_{11}=-g^x_2$, $g^{BCS}_{11}=g^{BCS}_4$, $g^c_{12}=2g^d_3-g^x_3$,
 $g^s_{12}=-g^x_3$, $g^{BCS}_{12}=g^{BCS}_3$.}.
 In the transition domain (where for example $0<|\v k_1+\v
 k_2|<\sqrt\Lambda$) the function is unknown, but, as we will argue
 shortly, its knowledge is not essential.

This parameterization of the coupling functions is a natural choice given the structure of the one-loop RG equations (\ref{1loop}) because the special couplings $g^{BCS}, g^x$ and $g^d$ get the strongest contributions from the PP, PH1 and PH2 diagrams, respectively. We say that $g^{BCS}$ is {\it 
resonant} in the PP channel, $g^x$ in the PH1 channel and $g^d$ in the PH2 channel. Furthermore the diagram with the largest contribution, say PP for a BCS coupling, again only includes BCS couplings and does not mix with non-BCS processes. 

For example, let $\v k\in P_1$ and $\v k'\in P_2$ such that $g(\v k,-\v k,-\v k',\v k')\approx g^{BCS}_3$ and consider the RG equation for this process:
\be 
\frac\ud{\ud\Lambda}g^{BCS}_3=\mbox{PP}+\mbox{PH1}+\mbox{PH2}\label{ex}
\ee
The dominant diagram PP, shown in Fig. \ref{exdiags}, involves the couplings $g^{BCS}(\v k,\v p)$ and $g^{BCS}(\v p,\v k')$. Within our approximation they are replaced by the constants $g^{BCS}_3$ or $g^{BCS}_4$, respectively. The contribution from diagram PP to Eq. (\ref{ex}) becomes
\be
\mbox{PP}=2B_{pp}^P(\Lambda,\v 0)\ g^{BCS}_3g^{BCS}_4,
\ee
where $B_{pp}^P(\Lambda,\v k)$ is the p-p bubble restricted to a saddle point patch of size $\rho$, given by Eq. (\ref{ppvH}) for $t'=0$. 

\begin{figure} 
        \centerline{\includegraphics[width=8cm]{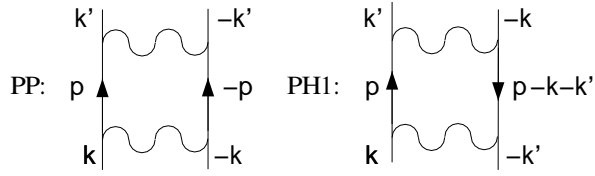}}
\caption{Two diagrams involved in Eq. (\ref{ex}). $\v k$ and $\v k'$ are typical vectors belonging to the patch $P_1$ and $P_2$, respectively; $\v p$ is the integration variable}\label{exdiags}\end{figure}

By contrast, the non-resonant diagram PH1 (also shown in Fig. \ref{exdiags}) involves couplings like $g(\v k,\v p-\v k-\v k',-\v k',\v p)$ which for almost every value of $\v p$ do not satisfy one of the special relations $\v k_1+\v k_2= O(\sqrt\Lambda)$, $\v k_3-\v k_2=\v Q+ O(\sqrt\Lambda)$ or $\v k_3-\v k_1=\v Q+ O(\sqrt\Lambda)$. This diagram therefore includes only the general couplings $g_2$ and $g_3$ and no special couplings $g^{BCS}, g^d$ or $g^x$. Thus  
\be
\mbox{PH1}=2B_{ph}^P(\Lambda,\v k+\v k')\ g_2 g_3,\label{phasespace}
\ee
where $B_{ph}^P(\Lambda,\v k)=-B_{pp}^P(\Lambda,\v k-\v Q)$. It follows that the RG flow depends on the ratio $B_{ph}^P(\Lambda,\v k+\v k')/B_{pp}^P(\Lambda,\v 0)$. In order to obtain a closed set of equations we replace this ratio by a constant. 

Let us first define 
\be
\alpha(\Lambda, \v k+ \v k'):=\frac{B_{pp}^P(\Lambda,\v k+\v k'-\v Q)}{B_{pp}^P(\Lambda,\v 0)}
\ee
which varies in principle between $0$ and $1$. Its value is $1$ if $\v k+\v k'=\v Q$, i.e. if the process to be renormalized is at the same time a $g^{BCS}$ and a $g^x$. Such processes exist of course, but they will not influence the RG equations in a relevant way. According to Eq. (\ref{p1}) a region in the Brillouin zone of width $\sim\sqrt\Lambda$ or smaller can be safely ignored within logarithmic precision. We thus assume that $\v k+\v k'-\v Q\sim\sqrt\Lambda$ or bigger. The biggest values are obtained if $\v k+\v k'$ is parallel to the Fermi surface $\v k+\v k'-\v Q=(\kappa,\kappa)$. For $\kappa\geq\sqrt\Lambda$ we get from Eq. (\ref{ppvH})
\be 
\alpha(\Lambda,\v k+\v k')\leq\frac{\log{\frac{4\rho^2}{\Lambda+2\rho\sqrt\Lambda}}}{\log{\frac{4\rho^2}{\Lambda}}}\ \MapRight{\Lambda\ll\rho^2}\ \frac12.
\ee
In the following we replace $\alpha(\Lambda,\v k+\v k')$ by a constant $\alpha\leq1/2$.  All the nearly resonant diagrams are treated in the same way, i.e. they include general couplings $g_1,\ldots g_4$ only and their amplitude is reduced with respect to the resonant ones by a factor $\alpha\leq1/2$.  

Our approximation scheme leads to a set of RG equations for the special couplings $g_1^d,g_2^x,g_3^{BCS},g_3^d,g_3^x,g_4^{BCS}$
\ba
\frac\ud{\ud l}g_1^d&=&2g_1^d(g_2^x-g_1^d)+2g_3^d(g_3^x-g_3^d)\nonumber\\
\frac\ud{\ud l}g_2^x&=&\left(g_2^x\right)^2+\left(g_3^x\right)^2\nonumber\\
\frac\ud{\ud l}g_3^{BCS}&=&-2g_3^{BCS}g_4^{BCS}+2\alpha\, g_3(2g_2-g_1)\nonumber\\
\frac\ud{\ud l}g_3^{x}&=&2g_2^{x}g_3^{x}+2\alpha\, g_3(g_2-g_1-g_4)\label{1stapp}\\
\frac\ud{\ud l}g_3^{d}&=&2(-2g_1^{d}g_3^{d}+g_2^{x}g_3^{d}+g_1^{d}g_3^{x})+2\alpha\, g_3(g_2    -g_4)\nonumber\\
\frac\ud{\ud l}g_4^{BCS}&=&-\left(g_3^{BCS}\right)^2-\left(g_4^{BCS}\right)^2.\nonumber
\ea

The general couplings $g_1,\ldots,g_4$ are resonant in none of the three channels. The RG flow of $g_1,\ldots,g_4$ is thus given by Eqs. (\ref{differential}), with the right hand side multiplied by $\alpha$.

Eqs. (\ref{1stapp}) can be rewritten in terms of the couplings which are associated with the dominant instabilities $g^s_\pm=-g^x_2\mp g^x_3,g^c_\pm=2g_1^d-g_2^x\pm(2g^d_3-g_3^x)$ and $g^{BCS}_\pm=g^{BCS}_4\pm g^{BCS}_3$
\ba 
\frac\ud{\ud l}g^s_\pm&=&-\left(g_\pm^s\right)^2\pm2\alpha\, g_3(g_1-g_2+g_4)\nonumber\\
\frac\ud{\ud l}g^c_\pm&=&-\left(g_\pm^c\right)^2\pm2\alpha\, g_3(g_1+g_2-g_4)\label{1stapp2}\\
\frac\ud{\ud l}g^{BCS}_\pm&=&-\left(g_\pm^{BCS}\right)^2\pm2\alpha\, g_3(2g_2-g_1).\nonumber
\ea
We see that for $\alpha=0$ it is a set of six independent equations, one for each instability. In fact, if the non-resonant diagrams are completely neglected, the RG becomes equivalent to the summation of ladder diagrams (see section \ref{consistency}). Even for $0<\alpha<1$ the special couplings associated with the different instabilities still do not influence each other, but each RG equation has a source term coming from the general couplings $g_1,\ldots,g_4$. For $\alpha=1$ and initial conditions  $g_1^d=g_1$, $g_2^x=g_2$, $g_3^d=g_3^x=g_3^{BCS}=g_3$, $g_4^{BCS}=g_4$, Eq. (\ref{differential}) is recovered (since these conditions are then conserved by the RG flow).

One can search for asymptotic solutions of the form $g(l)=\tilde g\cdot(l_c-l)^{-1}$ of Eq. (\ref{1stapp2}) by solving the resulting algebraic equations for the $\tilde g$. We first consider the possibility of diverging general couplings $g_1,\ldots,g_4$. In this case it follows from our analysis of Eq. (\ref{differential}) that the asymptotic behavior of the general couplings is given by $\tilde g_1=0,\ \tilde g_2=-\tilde g_4=1/(6\alpha)$ and $\tilde g_3=\pm\sqrt5/(6\alpha)$, depending on the sign of $g_3$. By inserting this behavior into Eq. (\ref{1stapp2}) it is easily seen that a real solution for $\tilde g^s_\pm,\tilde g^c_\pm$ and $\tilde g^{BCS}_\pm$ requires $\alpha\geq\sqrt{80/81}\approx0.994$. But as we argued above, the appropriate values of $\alpha$ are $\leq1/2$.

It follows that for acceptable values of $\alpha$ a special coupling can only diverge if $\tilde g_1=\ldots=\tilde g_4=0$. This means that this special coupling constant diverges at a higher energy scale than the general
couplings. 

The most striking difference to the ``one-dimensional'' solution is that one of the six couplings $g^\diamond_\pm$ can diverge, while all the others remain finite. This occurs here because the mixing of the flow for these couplings has been neglected on the basis of a phase space argument (see the discussion  before Eq. (\ref{phasespace})). This argument is certainly valid as long as the coupling function is slowly varying, but it may be questioned close to the instability, where the coupling function gets peaked.  It is argued however in Section \ref{consistency} that the non leading couplings can nevertheless stay finite in the case of an instability.

The divergence of the leading coupling is characterized by $\tilde g^\diamond_\pm=-1$. 
 Eq. (\ref{crit}) then implies
the asymptotic behavior $\chi\sim(\Lambda-\Lambda_c)^{-1}$,
corresponding to a mean field exponent. In view of 
Eq. (\ref{gkksol}) the coupling function $g^\diamond(\v k,\v k')$
diverges as $(\Lambda-\Lambda_c)^{-1}$ everywhere on the Fermi
surface and remains a smooth function upon renormalization, as anticipated in Section \ref{squareFS}. 

\subsection{Phase diagram}\label{phasediagram}

For a circular Fermi surface the dominant instability is superconductivity, as discussed in Section \ref{circ}. For the square Fermi surface the flow equations (\ref{1stapp2}) show that there are several possible instabilities, $s$- and $d$-wave superconductivity, charge and spin density waves, charge and spin flux phases. The values of the initial couplings will determine which of these instabilities, if any, occurs first, i.e. at the largest energy scale. To be specific we consider  an initial interaction consisting of 
on-site and nearest-neighbor terms 
\be 
\hat H_I=U\sum_{\v r}\hat n_{\up \v r}\hat n_{\down \v r}+V\sum_{\langle\v 
r,\v r'\rangle}\hat n_{\v r}\hat n_{\v r'}+J\sum_{\langle\v r,\v r'\rangle}
\hat{\v S}_{\v r}\hat{\v S}_{\v r'},
\ee
where $\hat n_{\v r}=\sum_\sigma \hat n_{\sigma \v r}$ and $\hat{\v S}_{\v r}$
 are the usual charge and spin operators on the lattice site $\v r$, the sum 
$\sum_{\langle\v r,\v r'\rangle}$ is over nearest neighbor bonds and $U$,$V$ 
and $J$ are parameters. The functional integral formulation of this model  is
 of the form (\ref{int}) with a coupling function 
\be
g(k_1,\ldots,k_4)=U-(V- J/4)e_{\v k_3-\v k_2}+J/2\, e_{\v k_3-\v k_1}
\label{gexample}
\ee
 where $e_{\v k}=-2(\cos k_x+\cos k_y)$.

The correct initial conditions for the flow equations would be the effective values of the couplings $g_1,\ldots,g_4$ at a cutoff 
$\Lambda_0\ll\rho^4$, when the flow enters the asymptotic regime. 
 It is at present impossible to connect in a controlled way these 
effective low energy couplings to the parameters of the microscopic
 interaction because this would imply solving Eq. (\ref{1loop}) without
 further expansion in powers of $\Lambda$. We therefore take the bare values of the couplings 
\ba
 g_1=g^d_1=U-4V-J,&\qquad g_2=g_2^x=U+4V+J,\nonumber\\
 g_3=g^d_3=g^x_3=g^{BCS}_3=U-4V+3J,&\qquad   g_4=g^{BCS}_4=U+4V-3J \label{UVJ}
\ea
as starting values instead of the (unknown) renormalized ones. This choice is a good approximation if $U$, $V$ and $J$ are small enough so that the couplings vary little before entering the asymptotic regime.

\begin{figure} 
        \centerline{\includegraphics[width=8cm]{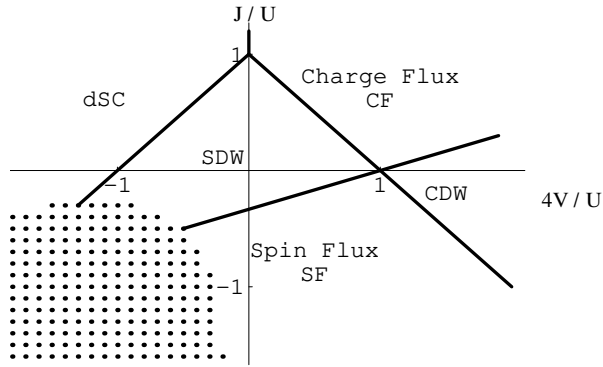}}
\caption{The phase diagram for $U>0$ and $\alpha=1/2$. In the dotted region all the couplings are flowing to zero. The size of the dotted region depends on the parameter $\alpha$, the other features are $\alpha$-independent}\label{upos}
\end{figure}
\begin{figure} 
        \centerline{\includegraphics[width=8cm]{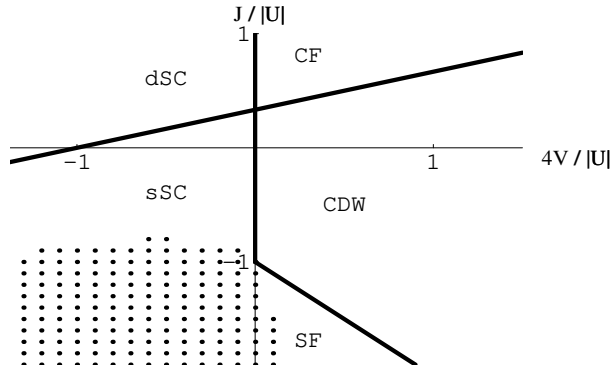}}
\caption{The phase diagram for $U<0$ and $\alpha=1/2$.}\label{uneg}  
\end{figure}

We have solved Eq. (\ref{1stapp2}) numerically for these initial conditions and obtained the phase diagrams of Figs. \ref{upos} and \ref{uneg} for, respectively, positive and negative values of $U$. The predicted phase for the repulsive Hubbard model ($U>V=J=0$) is a
 spin density wave (SDW) as expected and quite strong nearest neighbor
 terms are needed to establish a flux phase or (only for attractive 
$V$) a $d$-wave superconductor. An unexpected feature of Fig. 
\ref{upos} is, that the SDW can be destabilized by {\it 
positive} values of $J$.

In contrast to the ``old'' RG equations (\ref{differential}) our more elaborate scheme Eq. (\ref{1stapp}) produces only one diverging susceptibility while
 the others remain finite. Only at the phase boundaries, where the two
 neighboring phases are degenerate, both susceptibilities diverge.

In a certain parameter range (the dotted region in Figs \ref{upos} and \ref{uneg}) all the couplings are flowing to zero. In this case the behavior is not necessarily dominated by the saddle points and the RG flow has to be followed to order $\Lambda^{-1}$, including self energy corrections.

\subsection{A consistency test}\label{consistency}

The approximation scheme presented in section \ref{our} reduces to a
generalized random phase approximation (RPA) for $\alpha=0$. This approximation consists of summing a certain class of diagrams shown in Fig. \ref{ladders}. 

\begin{figure} 
        \centerline{\includegraphics[width=8cm]{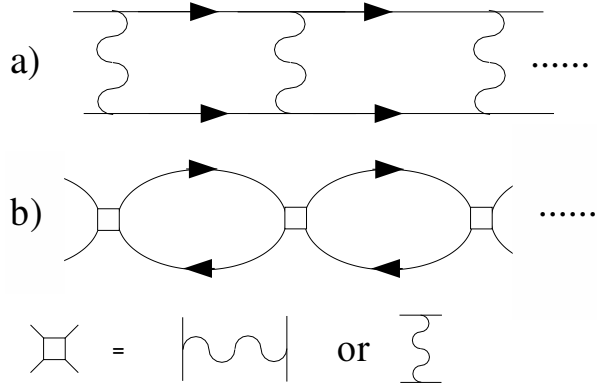}}
\caption{The class of diagrams summed in the generalized RPA: a) particle-particle ladders and b) the particle-hole ladders.}\label{ladders}  
\end{figure}

We have seen that although Eq. (\ref{1stapp}) at $\alpha>0$ goes beyond the RPA, its solutions are asymptotically the same as for $\alpha=0$. 
The reason is that in our approximate treatment of the non-resonant diagrams only the general couplings $g_1,\ldots,g_4$ intervene and not the (diverging) special couplings $g_1^d$, $g_2^x$ etc. It is not a priori clear whether this approximation is justified close to $\Lambda_c$. 

For example the (non-resonant) contribution of the diagram PH1 of Fig. \ref{exdiags} to the renormalization of $g_3^{BCS}$ involves an integration over the coupling function $g(\v k,\v p-\v k-\v k',-\v k',\v p)$, where $\v p$ is the integration variable moving along the one dimensional energy shell $|\xi_{\v p}|=\Lambda$. It is an integral over a one dimensional curve in the space of momenta $(\v k_1,\v k_2,\v k_3)$. For special values of $\v p$, the curve crosses one of the planes specified by Eq. (\ref{specialcouplings}). For example for $\v p=\v k'$ the coupling is equal to $g_3^{BCS}$. The question is whether one can neglect the contribution close to this point even when $g_3^{BCS}$ diverges.

In order to answer this question we have to know the coupling function
for small but finite total momentum. Within the generalized RPA one obtains
\be
g_\pm^{BCS}(\Lambda,\v q)=\frac{g_\pm^{BCS}(\Lambda_0)}{1+g_\pm^{BCS}(\Lambda_0)\int_{\Lambda}^{\Lambda_0} B^P_{pp}(\tilde\Lambda,\v q)\,\ud\tilde\Lambda},\label{gq}
\ee  
where $\v q$ is the (small) total momentum\footnote{$\v q=\v p-\v k'$ in the example of section \ref{our}.} and $g_\pm^{BCS}(\Lambda_0)$ is the ($\v q$- independent) initial value. Eq. (\ref{gq}) can be obtained either by solving the RG equation (\ref{1stapp2}) for $\alpha=0$ or by explicitly summing the ladder diagrams shown in Fig. \ref{ladders} a). A similar expression is obtained for $g^s=-g^x$ and $g^c=2g^d-g^x$ as functions of the deviation of the momentum transfer from $\v Q=(\pi,\pi)$. 

For negative values of $g_\pm^{BCS}(\Lambda_0)$ the effective coupling diverges at a
critical scale $\Lambda_c$. For $\Lambda>\Lambda_c$,  $g_\pm^{BCS}(\Lambda,\v q)$ has a maximum at $\v q=0$, which diverges for $\Lambda\to\Lambda_c$.

We have estimated the contribution of the peak in $g_\pm^{BCS}(\Lambda,\v q)$ to the non-resonant diagram by integrating this function over a curve in $\v q$-space. We find that such an integral diverges at most like $\log{(\Lambda-\Lambda_c)}$ and therefore is negligible compared to the resonant diagrams $\sim{(\Lambda-\Lambda_c)}^{-2}$. We conclude that the approximation presented in section \ref{our}, namely evaluating the non-resonant diagrams with the constant general values $g_1,\ldots,g_4$, is consistent.

The weak divergence of the non resonant diagrams are also consistent with the fact stated in Section \ref{our}, that non leading couplings and susceptibilities remain finite at $\Lambda_c$, since $\int_{\Lambda_0}^{\Lambda_c}\ud\Lambda\log(\Lambda-\Lambda_c)<\infty$. Note that this behavior can only be obtained because the momenta are allowed to move continuously on the Fermi surface. If instead we would discretize the Fermi surface and replace the continuous coupling function by a finite set of constants, a divergence of one coupling $g_c$ at a scale $\Lambda_c$ would imply the divergence, at the same scale $\Lambda_c$, of all these couplings that have $g_c$ appearing in the right hand side of the RG equation. 

Recently, the discretized RG has been studied in detail for a simpler model without van Hove singularities and without Umklapp scattering  \cite{Sebastien}.
It was found that there is a factor $1/N$ between the biggest non-dominant couplings and the dominant ones, where $N$ is the number of patches. This is consistent with our result in the continuous case ($N\to\infty$), that non-dominant couplings stay finite at $\Lambda_c$ while the dominant couplings diverge.
Similar results were also found in the large-$N$ limit of half-filled $N$-leg ladders \cite{Ledermann}.

\section{Special symmetries}\label{symmsection}

We will now discuss the phase diagram at half filling (Fig. \ref{upos} and
\ref{uneg}) in terms of special symmetries which turn out to be
present on the lines separating two different phases.
For that purpose it
is convenient to write the model in the Hamiltonian formalism 
\be\hat H=\hat H_0+\hat H_I-\hat H'\ee
with
$$\hat H_0=\sum_{\sigma,\v k}\xi_{\v k}\hat n_{\sigma \v k},$$
\ba 
\hat H_I&=&\frac12\frac1V\sum_{\v k_1\cdots \v k_4}
\delta_{\v k_1+\v k_2,\v k_3-\v k_4}\nonumber\\
& &\quad\times g(\v k_1, \v k_2, \v k_3, \v k_4) \sum_{\sigma,\sigma'}
\hat c^\dagger_{\sigma \v k_1}\hat c^\dagger_{\sigma' \v k_2}\hat c_{\sigma' \v
k_3}\hat c_{\sigma \v k_4}\nonumber
\ea
and
$$\hat H'=\frac12\sum_{\sigma,\v k}\left(\frac1V\sum_{\v p}\left(2g^{f}(\v k,\v p)-g^{e}(\v k,\v p)\right)
\right)\hat n_{\sigma \v k}.$$
The Hartree-Fock term $\hat H'$ has been included in order to keep $\hat H$ 
particle-hole symmetric (i.e. invariant with respect to the canonical
transformation $\hat c_{\sigma \v k}\to\hat c^\dagger_{\sigma \v k+\v Q}$).

The symmetry group of the noninteracting Hamiltonian $\hat H_0$ is
extremely large. From any function $d_{\v k}$ we can build an operator 
$N_d=1/2\sum_{\sigma,\v k}d_{\v k}\,\hat n_{\sigma \v k}$
that commutes with $\hat H_0$ and thus generates a continuous group of symmetry
transformations $\exp(i\alpha N_d)$. We find that $N_d$ commutes with the complete Hamiltonian $\hat H$ if and only if
\be 
\left(d_{\v k_1}+d_{\v k_2}-d_{\v k_3}-d_{\v k_4}\right)\, g(\v k_1,\ldots,\v
k_4)=0\quad \forall\ \v k_1,\ldots \v k_4.\label{condNd}
\ee

If $d$ is a suitably chosen $d$-wave function, such a symmetry can relate $s$- and $d$-wave superconducting order parameters by $[N_d,\hat
O_{sSC}]=-\hat O_{dSC}$ and $[N_d,\hat
O_{dSC}]=-\hat O_{sSC}$. In this situation on can transform $O_{sSC}$ into $O_{dSC}$ and vice versa by a symmetry operation. As a consequence the susceptibilities for $s$- and $d$-wave superconductivity must be exactly equal, provided condition (\ref{condNd}) holds. The symmetry $N_d$ relates in a similar way spin- or charge density waves to the corresponding flux phases. It might
therefore control the transition lines SDW/spin flux, CDW/charge flux and sSC/dSC. 

Similarly the operators 
${\vec S_d}=1/2\sum_{\sigma,\sigma',\v k}d_{\v k}\,\hat
c^\dagger_{\sigma \v k}\vec \tau_{\sigma\sigma'}\hat c_{\sigma' \v k}$
relate the spin density wave to the charge  flux
phase and the charge density wave to the spin flux
phase. They commute with $\hat H$ if the following two conditions hold 
\be 
\left\{\begin{array}{l}
(d_{\v k_1}- d_{\v k_2}+d_{\v k_3}-d_{\v k_4})\, g(\v k_1,\ldots,\v
k_4)=0\quad \forall\ \v k_1,\ldots \v k_4\\
(d_{\v k_1}+ d_{\v k_2}-d_{\v k_3}-d_{\v k_4})\, (1-X)g(\v k_1,\ldots,\v
k_4)=0\quad \forall\ \v k_1,\ldots \v k_4.
\end{array}\right.\label{condSd}
\ee

Another symmetry introduced by Lieb \cite{Lieb} and then further
 investigated by Yang and Zhang \cite{SO4} is generated by the 
pseudo spin operator $\eta_s=\sum_{\v k}s_{\v
k}\,\hat c_{\up\v Q-\v k}\hat c_{\down\v k}$, where the function
 $s_{\v k}$ satisfies $s_{\v Q-\v
k}=s_{\v k}$. It turns a $s$-wave  superconductor
into a charge density wave and  a $d$-wave superconductor into a charge flux phase (and vice
versa). $\eta_s$ commutes with $H_0$ because of the exact nesting
($\xi_{\v Q-\v k}+\xi_{\v k}=0$) and it commutes with the full
Hamiltonian provided
\be 
\left\{\begin{array}{l}
\sum_{\v p}s_{\v p}\,g(\v p,\bar\v p,\bar\v k,\v k)= s_{\v k}\sum_{\v
p}(2-X)g(\v k,\v p,\v p,\v k)\quad \forall\ \v k\\
s_{\v k_1}(1-X)g(\v k_1,\ldots,\v k_4)-s_{\v k_3}g(\bar\v k_3,\v
k_2,\v k_4,\bar\v k_1)+s_{\v k_4}g(\bar\v k_4,\v k_2,\v k_3,\bar\v
k_1)=0
\end{array}\right.\label{condeta}
\ee
$\forall\ \v k_1,\ldots \v k_4$, where $\bar\v k:=\v Q-\v k$.

Finally Zhang \cite{SO5}  considered the operators $\vec \Pi_d=1/2
\sum_{\sigma,\sigma',\v k}d_{\v k}\,\hat
c_{\sigma \v Q-\v k} \left(\vec\tau\tau^y\right)_{\sigma\sigma'}\hat
c_{\sigma' \v k}$ connecting a spin density wave to a $d$-wave superconductor and a spin flux
phase to a $s$-wave superconductor. The symmetry condition
is of the same form as (\ref{condeta}) but with
$s_{\v k_1}$ replaced by a function $d_{\v k_1}$ that satisfies  $d_{\v Q-\v
k}=-d_{\v k}$. 

It is in general difficult to satisfy the conditions (\ref{condNd}) to (\ref{condeta}). For example they do not hold for the $U-V-J$ interaction. The only exception is the pseudo spin
symmetry $\eta_s$ which is exact for $V=0$ and $s_{\v k}=1$. However the restriction
of the model to the two saddle point patches has more chance of being
symmetric. We take $s_{\v k}=1$ everywhere whereas $d(\v k)=1$ for $\v k
\in P_1$ and $d(\v k)=-1$
for $\v k\in P_2$. For this simple choice the symmetry
generators $N_d$, $\vec S_d$, $\eta_s$ and $\vec \Pi_d$ together with the total spin- and charge operators form a $so(6)\oplus so(2)$ Lie algebra. The commutation relations of the symmetry generators and the relevant order parameters
are listed in reference \cite{SO8}. 

We further assume that the (initial) coupling function $g(\v
k_1,\ldots,\v k_4)$ takes only four different values $g_1,\ldots g_4$ as in Eq. (\ref{g14}). The symmetry conditions are then $g_3=0$ for  $N_d$,  $g_1=0$ for
$\vec S_d$,  $g_2+g_4=2g_1$ for  $\eta_s$ and   $g_2+g_4=0$ for  $\vec
\Pi_d$. These hyper-planes in our four dimensional coupling space define exactly the transition planes of the phase diagram (shown in Figs. \ref{upos} and \ref{uneg} for $g_1,\ldots,g_4$ parametrized by $U,V$ and $J$).

We have thus shown that the
transition planes of the phase diagram are fixed by exact symmetries of the $g_1,\ldots,g_4$- model. This is a strong indication that the phase 
diagram shown in Figs. \ref{upos} and \ref{uneg} is the correct one at sufficiently weak coupling.
Such a determination of an exact phase diagram by simple 
symmetry considerations was also possible for a one-dimensional
 system \cite{Dionys}. 

In the more general situation of Section \ref{our} with ten instead of
four coupling constants the symmetry conditions read
\ba
g_3^d=g_3^x=g_3^{BCS}=g_3=0&\mbox{ for }& N_d,\nonumber\\
g_1=g_1^d=g_3^d-g_3^x=0&\mbox{ for }&\vec S_d, \nonumber\\
2g_1-g_2-g_4=2g_1^d-g_2^x-g^{BCS}_4=2g_3^d-g_3^x-g^{BCS}_3=0&\mbox{ for }&\eta_s, \\
g_2+g_4=g_2^x+g^{BCS}_4=g_3^x-g^{BCS}_3=0&\mbox{ for
}&\vec \Pi_d. \nonumber
\ea
 These
symmetries are respected by our approximate RG equation (\ref{1stapp})
for every value of the parameter $\alpha$. In fact it is
easy to show that if one of these conditions is satisfied at the initial scale $l_0$ it remains to
be so at any scale $l$. This is not completely trivial since an
arbitrary approximation scheme might violate the symmetries of the
model.

\section{Conclusion}\label{conclusion}

In summary, we have analyzed systematically the 
 instabilities of weakly interacting electrons with a square Fermi surface.
 Besides $s$- and $d$-wave superconductivity we have identified
 commensurate density waves and flux phases both in the charge 
and spin sector as the dominant instabilities. The transition lines of the phase diagram are fixed by exact symmetries and therefore robust for various approximation schemes. 

On a technical level, we have found that the dominant RG flow of the 
effective interaction in the limit of small energies is controlled
 by scattering processes with momenta close to van Hove points.
 Nevertheless, the low energy effective action contains relevant 
couplings between electrons everywhere near the Fermi surface. The
couplings far away from the saddle points diverge at the same critical
energy and with the same power $\sim(\Lambda-\Lambda_c)^{-1}$ as the
couplings at the saddle points. From this point of view, there is no
sign of a scenario with strong effective couplings at the
saddle points and weak couplings on the remaining Fermi surface.

We have shown that  the RG equations, although strongly coupled at the initial stage, become decoupled in the asymptotic limit of small energies.
Thus  the asymptotic result turns out to be similar to 
that of a generalized random phase approximation (RPA). The
 decoupling arises because the effective coupling function is strongly
 enhanced only for special configurations of the external momenta,
 i.e. in a small region of $\v k$-space. Thus the RG flow
 generates long ranged interactions in real space.

By contrast, in one dimension superconducting and density wave instabilities remain coupled. The 
low energy excitations of a one-dimensional electron gas are 
constrained to two privileged points in $\v k$-space: the Fermi 
points. This special geometry allows for a strong mixing between the
various interaction channels. The saddle points of the two-dimensional dispersion 
have a different status. They are privileged only due to the
 diverging density of states, but the low energy excitations
 exist on the whole Fermi surface. Therefore typical external momentum
configurations of an effective interaction are resonant in at most one channel.

Our analysis has further shown that a discretization of 
the Fermi surface in terms of a finite number of patches can 
enhance artificially the coupling between the different
 scattering channels in the low energy regime. In fact, in the numerical studies \cite{Drazen,Halboth,Honerkamp} different susceptibilities are found to diverge at a single energy
scale. By contrast, our results show that only the susceptibility of the dominant instability diverges. This
 behavior has been referred to as the ``moving pole solution'' by
the Russian school \cite{Dzyalo3}. 

The decoupling of the RG equations admittedly has only been
established to leading logarithmic order in the energy cutoff
$\Lambda$. This is justified for weak bare couplings for the
square Fermi surface. Subleading contributions have to be taken into
account if the Fermi surface is not nested at the van Hove filling or
if the (bare) interaction is not small enough. 

In order to estimate how small the bare interaction must be, we recall that our approach requires a patch around the saddle point small compared to the size of the Brillouin zone ($\rho\ll\pi/2$). On the other hand, the density of states of this patch has to be big compared to the remaining part of the Brillouin zone (in view of Eqs. (\ref{p1}) and (\ref{p2}) this means $\log4/\rho^2\ll\log4\rho^2/\Lambda$). The bare interaction $g_{\Lambda_0}\sim U$ must be small enough such that 
$$2\,U\int\!\ud\Lambda\, B^P_{pp}(\Lambda,\v 0)\approx \frac U{4\pi^2}\log^2{\frac{4\rho^2}\Lambda}\sim 1.$$
(The factor $2$ appears because there are two saddle point patches in the Brillouin zone.)  If for each of the two ``$\ll$'' signs above, a factor of 10 is introduced, this amounts to $U\sim0.02\,t$. A less stringent factor of 3 for each of the two inequalities would correspond to $U\sim0.6\, t$.      

{\bf Acknowledgments}
We are grateful to Drazen Zanchi and Alvaro Ferraz for many valuable 
discussions. B. B. also thanks F. Guinea for his helpful
 comments in the early stage of this work, the
 members of the Laboratoire de Physique Theorique et Hautes
 Energies in Jussieu (Paris) for their hospitality 
and Manfred Salmhofer for a useful information. 
This work has been carried 
out in the framework of a ``cotutelle de th\`ese'' between 
the universities of Fribourg and Paris 7. It was supported by
 the Swiss National Foundation through grants no. 53800.98 and 20-61470.00.


\end{document}